\documentclass[a4paper, 12pt]{article}
\usepackage{epsfig, psfrag}

\usepackage{afterpage}

\usepackage{amsfonts, amsmath, amssymb, amsthm}

\usepackage{tabls}
\usepackage{multirow}
\usepackage{rotating}

\usepackage{bbm}

\newcounter{thmNumber}[section]
\newcounter{lemmaNumber}[section]

\numberwithin{thmNumber}{section}
\numberwithin{lemmaNumber}{section}
\numberwithin{corollaryNumber}{section}
\numberwithin{definitionNumber}{section}
\newtheorem{theorem}[thmNumber]{Theorem}

\newtheorem{lemma}[lemmaNumber]{Lemma}

\newcommand{\R}{{\mathbb R}}
\newcommand{\C}{{\mathbb C}}
\newcommand{\Z}{{\mathbb Z}}
\newcommand{\N}{{\mathbb N}}

\newcommand{\be}{\begin{equation}}
\newcommand{\ee}{\end{equation}}

\newcommand{\Rhat}{\hat{R}}
\newcommand{\LL}{{\bf\underline{L}}}
\newcommand{\SSS}{{\bf\underline{S}}}

\newcommand{\Span}{\,\mbox{Span}\,}
\newcommand{\Conf}{\mathcal C}

\newcommand{\undZ}{\underline{Z}}

\begin{document}
\title{Asymptotics-based CI models for atoms:
properties, exact solution of a minimal model for Li to Ne, and application to atomic spectra}

\author{Gero Friesecke and Benjamin D. Goddard}

\date{February 25, 2009}

\maketitle

\begin{abstract} Configuration-Interaction (CI) models are approximations to the electronic Schr\"odinger
equation which are widely used for numerical electronic structure calculations in quantum chemistry.
Based on our recent closed-form asymptotic results for the full atomic Schr\"odinger equation in the limit
of fixed electron number and large nuclear charge \cite{PT}, we introduce a class of CI models for atoms
which reproduce, at fixed finite model dimension, the correct Schr\"odinger eigenvalues and eigenstates in this limit.

We solve exactly the
ensuing minimal model for the second period atoms, Li to Ne.
The energy levels and eigenstates are in remarkably good agreement with experimental data (comparable to that
of much larger scale numerical simulations in the literature), and facilitate a mathematical understanding
of various spectral, chemical and physical properties of small atoms.
\end{abstract}

\section{Introduction}
From the early days of quantum mechanics it has been clear that the chemical behaviour of atoms and molecules
is governed by their energy levels and electron configurations, which in turn are determined, to
very high accuracy, by the eigenvalues and eigenstates of the Schr\"{o}dinger equation $H\psi=E\psi$. But
80 years on, high-accuracy numerical computation of such data remains a largely unresolved challenge, even
for the smallest of systems such as a single Carbon atom. The only computations of which we are aware which meet the
mathematical ideal \cite{BLWW04} of
convergence tables showing an increasing number of converged digits
as a function of basis set size or number of iteration steps (for a reproducibly documented algorithm for the original
problem) concern two-electron systems such as He and H$_2$.
See \cite{Kurokawa08} for recent advances and references.

The underlying reasons are two-fold.

First, a ``curse of dimension'' phenomenon is present:
the Schr\"{o}\-dinger equation for an atom or molecule with $N$ electrons is a partial differential equation in $\R^{3N}$,
so direct discretization of each coordinate direction into $K$ gridpoints yields $K^{3N}$ gridpoints.
Thus the Schr\"odinger equation for a single Carbon atom ($N=6$) on a ten point grid in each direction ($K=10$)
already has a prohibitive $10^{18}$ degrees of freedom.

Second, one is dealing with a tough multiscale problem: chemical behaviour is not governed by total energies,
but by small energy differences between competing states. Even for very small systems, these are typically several orders of magnitude
smaller than total energies. For instance, as shown in the table below, the spectral gap between ground state
and first excited state of the second period atoms is less than 1~$\%$ of the total size of these energy
levels in all cases, and only about 0.1~$\%$ for Carbon, Nitrogen and Oxygen. Nevertheless this tiny gap
is of crucial importance, as the two states it separates have different spin and angular momentum symmetry, and hence
completely different chemical behaviour.

\begin{table}[ht]
\resizebox{\textwidth}{!}{
  \begin{tabular}{|c|c|c|c|c|c|c|c|c|c|c|}
   \hline
   Atom  & Li & Be  & B  & C  & N  & O  & F & Ne  \\
   \hline
   $\begin{array}{l}\mbox{\small Ratio of first spectral gap}\\
                    \mbox{\small to ground state energy}\end{array}$
         & 0.0093 & 0.0068  & 0.0053 & 0.0012  & 0.0016 & 0.00096 & 0.0078 & 0.0047 \\
   \hline
  \end{tabular}
} 
\vspace*{-2mm}
\caption{Multiscale structure of atomic spectra (experiment, \cite{NIST}).}
 \label{Tab:Multiscale}
\end{table}

To deal with the curse of dimension, in quantum chemistry a large array of reduced models has been developed. For small systems
with up to one or two dozen electrons, the most accurate and most widely used class of models are the Configuration-Interaction
(CI) models, whose origins go back to the early years of quantum
mechanics (see e.g. \cite{Hylleraas}) and whose systematic development started with the work of Boys \cite{Boys}
and L\"owdin \cite{Loewdin}. Roughly speaking, these are ``tensor product Galerkin approximations'':
the full electronic Schr\"odinger equation is projected onto a
subspace spanned by carefully chosen Slater determinants
(= antisymmetrized tensor products), which are in turn formed from a small set of
orbitals (= elements of the single-particle Hilbert space $L^2(\R^3\times\Z_2)$).

Different CI models differ
by the choice of orbitals and the selection of the subset of Slater determinants. The question of how to best
make these choices remains the subject of a great deal of current research in the quantum chemistry literature, with
the best methods to date relying on a combination of chemical intuition, computational experience, and nonlinear
parameter optimization, as well as on a huge number (between 10$^6$ and 10$^9$) of included determinants.
See \cite{SzaboOstlund96, HJO00} for a general overview of the CI method and its most common variants
such as Doubly excited CI (DCI), Multi-determinant Hartree-Fock (MDHF),
Complete active space self-consistent field method (CASSCF), Coupled-Cluster theory (CC), and the (desirable but usually
not practical) Full CI (FCI),
and see e.g. \cite{Tayloretal1, Tatewaki94, Komasa02, Bendazzoli04, CanalNeto05, NIST05, Nakatsuji07, Kurokawa08} for applications to atomic energy level calculations.

Our goal in this paper is to introduce, analyze, and apply to atomic energy level prediction
a particular class of CI models for atoms which exploit our recent closed-form asymptotic results for
the full atomic Schr\"odinger equation in the limit of fixed electron number $N$ and large nuclear charge $Z$ \cite{PT}.
Namely, we require that the model of fixed finite subspace dimension $K$ reproduce correctly the first $K$
Schr\"odinger eigenvalues and eigenstates in this limit.

That such a requirement can be met by a fixed-resolution CI model is not trivial (for example, it is not met by
Hartree-Fock theory, even in an infinite, complete one-electron basis), but a simple consequence of
the asymptotic results in \cite{PT} (see Section \ref{Sec:Asy}).

The above limit exhibits the important
multiscale effect that the ratio shown experimentally in Table \ref{Tab:Multiscale} of first
spectral gap $\Delta E$ to ground state
energy $E$ of the Schr\"odinger equation tends to zero \cite{PT}.
The requirement that the corresponding eigenstates and gaps be nevertheless captured correctly by an approximation
should hence be relevant to yielding good eigenstates and gaps in the realistic situation when this ratio is small.

\begin{figure}[ht]
  \begin{center}
   \resizebox{0.45\textwidth}{!}{
    \includegraphics{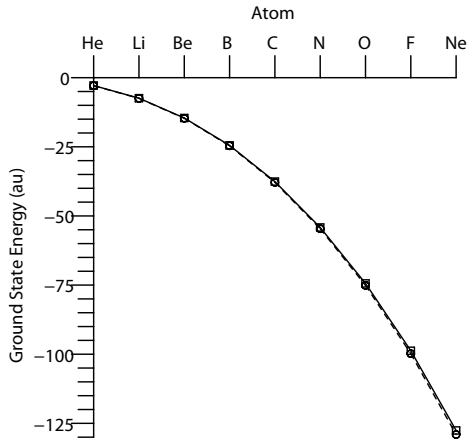}
   }
  \end{center}
  \caption{Experimental (dashed) and minimal asymptotics-based CI (solid) ground state energies of atoms.}
  \label{Fig:GSEnergies}
\end{figure}

In fact, even the minimal asymptotically correct CI model for atoms and ions with 1 to 10 electrons (eq. (A'), (B'),
(C') in Section \ref{Sec:Asy}), whose subspace dimension turns out to be 8, 28, 56, 70, 56, 28, 8 for Li, Be, B, C, N, F,
turns out to be very interesting.
\begin{itemize}
\item[(i)] We find that the requirement of asymptotic correctness leads to
Slater orbitals $P(x)e^{-\gamma|x|}$ (where $P$ is a polynomial
and $\gamma>0$ a constant), not Gaussian orbitals $P(x)e^{-\gamma|x|^2}$ used in the overwhelming majority of
numerical CI computations on account of their easy facilitation of two-centre integral evaluation. See Section \ref{Sec:Asy}.
\item[(ii)] The model is exactly soluble (via the methods introduced in \cite{PT}), except for the determination of the
dilation parameters $\gamma$ in the Slater orbitals, which are straightforward to calculate numerically with very
high accuracy. See Section \ref{Sec:Sol}.
\item[(iii)] The model does remarkably well when compared to experimental data and high-dimensional simulations in the literature.
It captures around 99 percent of the ground state energy in all cases, without a single empirical parameter! See Figure
\ref{Fig:GSEnergies}.
Moreover the predicted ground state spin and angular momentum quantum numbers ($^1$S for He, Be, Ne, $^2$S for H and Li, $^4$S for N, $^2$P for
B and F, and $^3$P for C and O) come out right in each case; spectral gaps are captured well (and in a
significant number of cases more accurately than in the benchmark numerical multi-determinant
Hartree-Fock calculations of Tatewaki et al \cite{Tatewaki94} which used a much larger basis set); and for $N\ge 5$ the model
is never outperformed by more than a factor ten by any method, including large-scale simulations with subspace
dimension bigger than $10^7$. For a detailed comparison see Section \ref{Sec:Comparison}.
\end{itemize}

\noindent
Thus, our work yields for the first time few-parameter, explicit, closed-form approximations to
the low-lying eigenstates of the atoms Li, Be, B, C, N, O, F, Ne which are of chemically relevant accuracy. These
provide a hopefully useful reference for the calibration of numerical methods, and a valuable tool
to advance mathematical understanding of physical, chemical and spectral differences between the elements.
For example, the ground state wavefunctions confirm the basic mathematical picture of the periodic table obtained
in \cite{PT} by asymptotic analysis of the Schr\"odinger equation for strongly positive ions, and make it quantitative
for neutral atoms;
and they allow to trace the size of spectral gaps to individual Coulomb and exchange
integrals, thereby making more rigorous longstanding insights by quantum chemists and revealing the
cancellations that lead to the small size of gaps compared to total energies (Table \ref{Tab:Multiscale}).
See Section \ref{Sec:Comparison}.

Nevertheless a great many open problems remain, even for minimal asymp\-totics-based CI.

1) In this paper we demonstrate its accuracy via comparing to experimental values (see
Section \ref{Sec:Comparison}) and proving desirable theoretical properties (see Section \ref{sec:Asy}),
but how can it be understood in terms of rigorous error estimates comparing it to the Schr\"odinger equation?

2) In particular, why is the use of just one dilation parameter per one-electron subspace $V_j$
so effective? As far as we are aware, although screening parameters are widely used (see
Section \ref{sec:Asy}), there are no rigorous mathematical results regarding their
effectiveness. For instance, one might hope such parameters to emerge in some order expansion of
a suitably scaled problem.

3) How does the model fare for larger atoms? For this step we would suggest
automation of the calculation of the eigenspaces and energy expressions (analogous
to Tables \ref{Tab:VeeMatrix1}--\ref{Tab:VeeMatrix2}), and Fourier transforms and
one- and two-body integrals (as in Lemmas \ref{L:FT} and \ref{L:VeeElements}). We would hope the model
to show interesting chemical effects such as the shell ordering $4s<3d$, and its
occasional reversal, in the transition metals.

Finally, it is highly desirable that the asymptotics-based CI approach introduced
here be extended to molecular problems. The principal observation (Theorem \ref{T:Properties} (ii)) that CI models
of fixed finite subspace dimension can be constructed which
reproduce correctly the first K Schr\"odinger eigenvalues
in a large nuclear charge limit is not limited to atoms, as will be discussed elsewhere. But in the molecular case the
ensuing orbitals are not available in closed form, and hence do not lead so readily to a mathematical picture
of basic physical and chemical properties.
\section{Asymptotics-based CI models for atoms} \label{sec:Asy}
\subsection{General CI models}
We begin with a mathematical description of CI methods. We find it convenient
to do so in the more abstract setting of subspaces and subspace projections
rather than the, equivalent, setting of basis sets and expansion coefficients used in
the chemistry literature \cite{SzaboOstlund96}. Moreover we introduce a rigorous distinction between general and
symmetry-preserving CI methods. (Both of these, as well as hybrid methods in which the solution to a
non-symmetry-preserving model is projected a posteriori onto an invariant subspace, are in use in the chemistry literature.)

Starting point for the derivation of any CI model is
the exact (nonrelativistic, Born-Oppenheimer) time-independent Schr\"odinger equation
for atoms and ions which one seeks to approximate,
\be \label{SE}
    H \Psi = E \Psi,
\ee
where, for nuclear charge $Z>0$ and $N$ electrons and in atomic units,
\be \label{ham}
    H = \underbrace{\sum_{i=1}^N \Big( -\frac{1}{2} \Delta_{x_i} - \frac{Z}{|x_i|}
    \Big)}_{=:H_0} + \underbrace{\sum_{1 \leq i < j \leq N} \frac{1}{|x_i-x_j|}}_{=:V_{ee}},
\ee
$E\in\R$, and
\be \label{space}
    \Psi \in L^2_a\big((\R^3 \times \Z_2)^N\big).
\ee
Here and below the $x_i\in\R^3$ are electronic position coordinates, $s_i\in\Z_2=\{\pm\frac12\}$ are spin coordinates,
and $L^2_a$ is the usual Hilbert space of $N$-electron
functions $\Psi\, : \, (\R^3\times\Z_2)^N\to\C$ which are square-integrable,
\be \label{norm}
   \int_{\R^{3N}}\sum_{(\Z_2)^N} |\Psi(x_1,s_1,\dots,x_N,s_N)|^2 = ||\Psi||^2 <\infty,
\ee
and satisfy the antisymmetry principle that, for all $i$ and $j$,
\be \label{anti}
    \Psi(\dots, x_i,s_i, \dots, x_j,s_j, \dots)= -\Psi(\dots, x_j,s_j, \dots, x_i,s_i, \dots).
\ee
Mathematically, $H$ is a bounded below, self-adjoint operator with domain $L^2_a\cap H^2$, where
$H^2$ is the usual Sobolev space of $L^2$ functions with second weak derivatives belonging to $L^2$. It is known that
for neutral atoms ($Z=N$) and positive ions ($Z>N$), there exists an infinite number of discrete eigenvalues, the
corresponding eigenspaces being finite-dimensional (Zhislin's theorem, see \cite{Friesecke03} for a short proof).
\\[2mm]
Translating \cite{SzaboOstlund96} into mathematical terminology, a CI model is a tensor product Galerkin
approximation to the many-electron Schr\"odinger equation. More precisely:
\\[2mm]
{\bf Definition 2.1} A CI model of an $N$-electron system with Hamiltonian $H$ is a projection of the Schr\"odinger
equation (\ref{SE}) of form
\be \label{CIeq}
    PHP\Psi=E\Psi, \;\;\; \Psi\in V, \; \mbox{$V$ a subspace of $L^2_a$}, \; P=\mbox{projector onto
$V$},
\ee
with the additional requirement that $V$ must possess a basis consisting of Slater determinants.
\\[2mm]
Recall that a Slater determinant is an anti-symmetrized tensor product
$|\eta_1\dots\eta_N\rangle\in L^2((\R^3\times\Z_2)^N)$ of orthonormal one-electron functions $\eta_i\in L^2(\R^3\times\Z_2)$,
the antisymmetrization being necessary to comply with the quantum mechanical law
(\ref{anti}). The difference between different CI models lies in the freedom to choose the subspace $V$, or -- in quantum chemistry
language -- to select a set of orbitals and a set of Slater determinants to be included into the CI expansion.

Note that
if $V$ is spanned by the orthonormal Slater determinants $\Psi_\alpha$, $\alpha=1,\dots,M$, the projection operator $P$
onto $V$ has the expansion $\sum_{\alpha=1}^M|\Psi_\alpha\rangle\langle\Psi_\alpha|$, and eq. (\ref{CIeq}) can be written
in its more standard matrix form $Ac=Ec$, where $A$ is the M$\times$M matrix with entries
$A_{\alpha\beta}=\langle\Psi_\alpha| H | \Psi_\beta\rangle$, and $c$ is the coefficient vector in the expansion $\Psi=\sum_\alpha
c_\alpha\Psi_\alpha$. The more abstract form (\ref{CIeq}) emphasizes the elementary fact
that the CI eigenvalues and eigenstates only depend on the subspace $V$, not on the choice of basis within
this subspace.
\\[2mm]
A basic desirable feature of CI models, not related to the tensor product structure but only
to that of a linear subspace projection, is the following.
\begin{lemma} \label{L:UB} (Rigorous upper bound) For $j=1,\dots,\mbox{dim}\,V$, we have $E_j\le E^{CI}_j$, where
$E_j$ and $E_j^{CI}$ are the $j^{th}$ eigenvalue, including multiplicity, of the
Schr\"odinger equation (\ref{SE}) respectively the CI equation (\ref{CIeq}).
\end{lemma}
\noindent
{\bf Proof} This is an immediate consequence of the min-max theorem for discrete eigenvalues of
a self-adjoint operator below the bottom of the essential spectrum \cite{ReedSimon78}.
\subsection{A mathematical definition of the notion of configuration for atoms}
In the quantum chemistry literature, the word ``configuration'' is often employed as a synonym for
Slater determinant \cite{SzaboOstlund96}. But
in the atomic physics and atomic spectroscopy literature (e.g. \cite{NIST}), as well as some of the best
computational studies, the word
``configuration'' has a more subtle meaning, which takes into account the important role
played by spin and angular momentum symmetries. For our mathematical purposes, the latter notion turns out to be
very useful, so let us formalize it mathematically.

First, recall the total angular momentum operator $\LL=(L_1,L_2,L_3)$,
the total spin operator $\SSS=(S_1,S_2,S_3)$ and the parity operator $\hat{R}$, along with
the fact that the operators
\be
    \LL^2, \SSS^2, L_3, S_3 \text{ and } \Rhat \label{ops}
\ee
commute with each other and with $H$ (see \cite{PT} for the result, as well as a mathematical definition of the operators
$\LL$, $\SSS$ and (\ref{ops})).\footnote{On single-electron functions $\chi(x,s)$, $x\in\R^3$, $s\in\{\pm 1/2\}$, one has
$\LL=x\wedge\frac{1}{i}\nabla$, $(\hat{R}\chi)(x,s)=\chi(-x,s)$, and $S_\alpha$ is multiplication by a Pauli matrix,
$$
   \left(\!\!\begin{array}{c} (S_\alpha\chi)(x,1/2) \\ (S_\alpha\chi)(x,-1/2)\end{array}\!\!\right)
   \!=\! \sigma_\alpha
   \left(\!\!\begin{array}{c} \chi(x,1/2) \\ \chi(x,-1/2)
   \end{array}\!\!\right), \;
    \sigma_1:=\frac12\begin{pmatrix} 0 & 1 \\ 1 & 0 \end{pmatrix}, \;
    \sigma_2:=\frac12\begin{pmatrix} 0 & -i \\ i & 0 \end{pmatrix}, \;
    \sigma_3:=\frac12\begin{pmatrix} 1 & 0 \\ 0 & -1 \end{pmatrix}.
$$
} 

One starts from a finite number of mutually orthogonal subspaces of the single-electron Hilbert space,
\be \label{FCIonesub}
   V_1,...,V_k\subset L^2(\R^3\times\Z_2),
\ee
which are {\it irreducible representation spaces for the joint spin and angular momentum algebra
$Span\,\{L_1,L_2,L_3,S_1,S_2,S_3\}$}. In elementary terms,
this means that the subspaces must be of ``fixed angular and spin symmetry'' and ``minimal dimension'', more precisely:
each $V_i$ must be invariant under the $L_i$ and $S_i$, the operators
$\LL^2$ and $\SSS^2$ must be constant on $V_i$, and $V_i$ must have minimal dimension
(i.e. dimension $(2s+1)(2\ell+1)|_{s=1/2}=2\cdot (2\ell + 1)$ when $\LL^2=\ell(\ell+1)I$ and $\SSS^2=s(s+1)I$; note that the spin quantum
number $s$ equals $1/2$ for any $V_i$, since $\SSS^2=\frac34 I$ on the whole single-electron state space $L^2(\R^3\times\Z_2)$).
\\[2mm]
{\bf Definition 2.2} A {\it configuration} of an $N$-electron atom or ion is a subspace of N-electron state space (\ref{space}) of the following form:
\be \label{Conf}
  \Conf^{d_1,..,d_k} = \Span\Bigl\{ |\chi_1,\dots,\chi_N\rangle \, \Bigl|\Bigr. \, \{\chi_1,\ldots, \chi_N\} \mbox{ any ON set with }
          \sharp\{i\, | \, \chi_i\in V_j\}=d_j\Bigr\},
\ee
where $V_1,..,V_k\subset L^2(\R^3\times\Z_2)$ are mutually orthogonal
irreducible representation spaces of the joint spin and angular momentum algebra, and
$d_1,...,d_k$ is a partition of $N$ (i.e. $d_j\in\N\cup\{0\}$, $\sum_jd_j=N$).
\\[2mm]
The main point here is that {\it all} choices of the $\chi_i$'s consistent with the
requirement that a fixed number of them have to be picked from each $V_j$ have to be included.

As an elementary but important consequence, each configuration
is invariant under the spin and angular momentum operators $\LL$ and $\SSS$, and in particular under the operators (\ref{ops}). This
is immediate from the invariance of the $V_i$ under $\LL$ and $\SSS$ and the following identity for the application of one-body
operators to Slater determinants: $\sum_{j=1}^N A(j)|\chi_1\cdots\chi_N\rangle = \sum_{j=1}^N|\chi_1\cdots(A\chi_j)\cdots\chi_N\rangle$.
\\[2mm]
{\bf Example 1: The configurations $1s^22s$ and $1s^22p$ for Lithium.} \\
Let
\begin{align}
  V_1&=\Span\{\phi_{1s} \! \uparrow , \, \phi_{1s} \! \downarrow\}, \;\;\;
  V_2 = \Span\{\phi_{2s} \! \uparrow, \,\phi_{2s} \! \downarrow\},
  \notag \\
  V_3 &= \Span\{\phi_{2p_i} \! \uparrow , \, \phi_{2p_i} \! \downarrow \; | \; i=1,2,3 \},
  \label{examples1}
\end{align}
where the $\phi$'s are the hydrogen-like orbitals
\begin{align}
    \phi_{1s}(x)&:=\frac{Z_1^{3/2}}{\sqrt{\pi}} e^{-Z_1|x|} \notag\\
    \phi_{2s}(x)&:= \left(
    \tfrac{3Z_2^5}{8\pi(4Z_1^2-2Z_1Z_2+Z_2^2)} \right)^{1/2}
    \left(1- \tfrac{1}{6}(2Z_1 + Z_2)|x| \right)
    e^{-Z_2|x|/2} \label{CIorbitals}\\
    \phi_{2p_i}(x)&:=
    \frac{Z_3^{5/2}}{\sqrt{32 \pi}}x_i e^{-Z_3|x|/2}, \; i=1,2,3, \notag
\end{align}
$Z_1$, $Z_2$, $Z_3$ are positive parameters,
and $\uparrow$, $\downarrow$ denote the spin functions $\uparrow(s)=\delta_{\frac12}(s)$,
$\downarrow(s)=\delta_{-\frac12}(s)$.

Note that the orbitals in (\ref{examples1}) are orthonormal (hence the coefficients in $\phi_{2s}$),
and that for $Z_1=Z_2=Z_3=Z$ they reduce to
the standard eigenstates of the hydrogen atom Hamiltonian $-\frac12\Delta-\frac{Z}{|x|}$. The $Z_i$ will
play an important role later.

For $N=3$, choosing the partitions $d_1=2$, $d_2=1$, $d_3=0$ respectively $d_1=2$, $d_2=0$, $d_3=1$ yields the subspaces (or configurations)
\begin{align}
   \Conf^{2,1,0}&=\Span\{|\phi_{1s} \! \uparrow,\, \phi_{1s} \! \downarrow, \, \phi_{2s} \! \uparrow \rangle, \;
   |\phi_{1s} \! \uparrow,\, \phi_{1s} \! \downarrow, \, \phi_{2s} \! \downarrow \rangle\}, \nonumber \\
   \Conf^{2,0,1}&=\Span\{|\phi_{1s} \! \uparrow,\, \phi_{1s} \! \downarrow, \, \phi_{2p_i} \! \uparrow\rangle,
   \; |\phi_{1s} \! \uparrow,\, \phi_{1s} \! \downarrow, \, \phi_{2p_i} \! \downarrow \rangle\; | \; i=1,2,3\}.
   \label{Confdef}
\end{align}
We call these subspaces 1s$^2$2s$^1$ and 1s$^2$2p$^1$.
In chemistry this terminology is common to describe the structure of {\it individual wavefunctions},
but in the setting just introduced, it is independent of which wavefunction is chosen.

For $Z_1=Z_2=Z_3=Z$, these subspaces have the interesting physical meaning that they are the bottom two eigenspaces
of the Lithium atom Hamiltonian in first order perturbation theory \cite{PT}.
\\[2mm]
{\bf Example 2: The subspace $2p_1^12p_2^1$ for Helium} \\
The subspace
$$
   \tilde{C}=\Span\{|\phi_{2p_1}\!\uparrow,\phi_{2p_2}\!\downarrow\rangle, \; |\phi_{2p_1}\!\downarrow,\phi_{2p_2}\!\uparrow\rangle,
                    |\phi_{2p_1}\!\uparrow,\phi_{2p_2}\!\uparrow\rangle, \; |\phi_{2p_1}\!\downarrow,\phi_{2p_2}\!\downarrow\rangle \}
$$
is {\it not} a configuration, because the selection of Slater determinants does not correspond to the rule in Definition 2.1.
Indeed this subspace is not invariant under the spin and angular momentum algebra. For instance, applying
$L_3$ to the first Slater determinant gives $i(|\phi_{2p_1}\!\uparrow \phi_{2p_1}\!\downarrow\rangle -|\phi_{2p_2}\!
\uparrow \phi_{2p_2}\!\downarrow\rangle)$,
which lies outside the subspace.
\subsection{Symmetry-preserving CI models}
A general class of symmetry-preserving CI models can now be defined mathematically. We remark that
the principle of symmetry-preserving
numerical schemes has proved very successful in other areas of scientific computing, a prime example being symplectic schemes in
Hamiltonian dynamics \cite{LeimkuhlerReich}.
\\[2mm]
{\bf Definition 2.3} A symmetry-preserving CI model for an $N$-electron atom or ion with Hamiltonian $H$
is a finite-dimensional projection of the Schr\"odinger equation (\ref{SE}),
\be \label{FCIeq}
    PHP\Psi = E\Psi, \;\;\; \Psi\in V, \;\;\; \mbox{$V$ a subspace of $L^2_a$},\; \mbox{$P$=orthogonal projector onto }V,
\ee
with the additional requirement that
\be \label{FCIspace}
    V = \Span\{\Conf^{(1)},\dots,\Conf^{(K)}\},
\ee
where $V_1,..,V_k\subset L^2(\R^3\times\Z_2)$ is a collection of mutually orthogonal irreducible representation spaces
of the spin and angular momentum algebra, and each $\Conf^{(j)}$ is a configuration with respect to the $V_i$.
\\[2mm]
{\bf Example} Taking $V = \Span\{\Conf^{2,1,0},\,\Conf^{2,0,1}\} = \Span\{1s^22s, \, 1s^22p\}$ (with notation as
in Example 1) yields an invariant CI model for Lithium.
\\[2mm]
The fundamental point of Definition 2.3 is that unlike general CI, symmetry-preserving CI
retains the spin and angular momentum symmetries
of the atomic Schr\"odinger equation. In particular, eigenspaces retain well defined spin and angular momentum
quantum numbers $S$ and $L$ (see \cite{PT} for their mathematical definition):
\begin{lemma} \label{L:FCIsymm} (Symmetry preservation) For arbitrary $N$ and $Z$, and any invariant CI model (\ref{FCIeq}), (\ref{FCIspace}) as in Definition 2.3,
The operators (\ref{ops}) \\
(i) leave $V$ invariant \\
(ii) commute with the CI Hamiltonian $PHP \, : \, V \to V$. \\
In particular, any eigenspace of an invariant CI model possesses a basis of joint eigenstates of the operators (\ref{ops}).
\end{lemma}
\noindent
{\bf Proof} This is an elementary consequence of the invariance of individual configurations under $\LL$, $\SSS$ and $\hat{R}$.
Note that the underlying one-electron subspaces $V_i$, being $\LL^2$ eigenspaces with some eigenvalue $\ell(\ell+1)$, are automatically $\hat{R}$ eigenspaces,
with eigenvalue $(-1)^{\ell-1}$.
\\[2mm]
We note the well known fact that
the physically important property (i) is violated by standard approximations such as the Hartree-Fock approximation,
even when the individual orbitals have well defined spin and angular momentum quantum numbers. For instance, the
Slater determinant $|\phi_{2p_1}\!\uparrow \phi_{2p_2}\!\downarrow\rangle$
is neither an $\LL^2$ nor an $\SSS^2$ eigenstate.
\subsection{Asymptotics-based subspace selection} \label{Sec:Asy}
We now come to the, in applications crucial, issue of selecting a ``good'' CI subspace $V$ in the approximation (\ref{CIeq}).

Commonly, this relies on a great amount of chemical intuition, computational experience, and nonlinear optimization.
For example, one would employ the set of Slater determinants formed from the first $N+k$ eigenstates of the nonlinear Hartree-Fock equations of the
system under consideration (``k-fold excited CI''), solved numerically in a background subspace of dimension $L>>N+k$
spanned by Gaussian orbitals. For more information, common variants and refinements see \cite{SzaboOstlund96, HJO00}.

We propose here an alternative strategy, in which the intermediate step
of a Hartree-Fock calculation no longer appears, and which is based on three reasonable
theoretical requirements. The CI model should
\begin{itemize}
\vspace*{-2mm}
\item[1.] preserve the symmetry of the atomic Sch\"odinger equation under spatial and spin
rotation (see Definition 2.3)
\vspace*{-2mm}
\item[2.] preserve the virial theorem, i.e. eigenstates should have
the correct virial ratio of $-2$ between potential and kinetic energy
\vspace*{-2mm}
\item[3.] be asymptotically correct in the iso-electronic limit $Z\to\infty$.
\end{itemize}
By 3. we mean that the model (if its dimension is $K$) reproduces correctly the first $K$
Schr\"odinger eigenvalues and eigenstates in this limit (see Theorem \ref{T:Properties} for a precise statement).
Note that the limit of large $Z$ captures the
physical environment of inner shell electrons in large atoms. Also, recall its important theoretical feature that
the ratio of first spectral gap to ground state energy of the Schr\"odinger equation (\ref{SE})--(\ref{space}) tends to zero \cite{PT},
the experimental ratio for true atoms being very close to zero (see Table \ref{Tab:Multiscale}).

We now apply requirements 1, 2, 3 to the atoms Li to Ne, by not designing a largest such model which can be
handled computationally, but a {\it minimum-dimensional} model.
Below, $(Z_1,Z_2,Z_3)=:\undZ$ denotes the vector
of dilation parameters appearing in the orbitals (\ref{CIorbitals}). We first discuss the case of the ground state. The ensuing
{\bf minimal CI model for Li, Be, B, C, N, O, F, Ne ground states} is then:

\begin{table}[ht]
\resizebox{\textwidth}{!}{
  \begin{tabular}{|l l|}
   \hline
   (A) & (Choice of a parametrized, asymptotically exact family of subspaces) \\
       & $V(\undZ):= \mbox{Span} \{\Conf^{2,0,N-2},\Conf^{2,1,N-3},\Conf^{2,2,N-4}\}$, \\
       & with the $\Conf$'s as in (\ref{Conf}) and $V_1,V_2,V_3$ as in (\ref{examples1}), (\ref{CIorbitals}), \\[2mm]
   (B) & (Subspace eigenvalue problem) \\
       & $E^{CI}(\undZ) := \mbox{lowest eigenvalue of }P(\undZ)HP(\undZ)\mbox{ on }V(\undZ)$, \\
       & $\Psi^{CI}(\undZ) := \mbox{corresponding normalized eigenstate}$, \\
       & $\mbox{where }P(\undZ)=\mbox{orthogonal projector of }L^2_a((\R^3\times\Z_2)^N)\mbox{ onto }V(\undZ)$ \\[2mm]
   (C) & (Variational parameter determination) \\
       & $\undZ_* := \mbox{argmin}_{\undZ} E^{CI}(\undZ), \; E^{CI} := E^{CI}(\undZ_*), \; \Psi^{CI} := \Psi^{CI}(\undZ_*)$ \\[2mm]
   \hline
  \end{tabular}
} 
 \vspace*{-2mm}
 \label{Tab:MinimalCI}
\end{table}

(In (A), it is understood that only configurations $\Conf^{d_1,d_2,d_3}$ for which each $d_i$ is $\ge 0$ are included.)

Some remarks are in order.

(1) This model is certainly not the only conceivable model which satisfies 1., 2., 3.,
especially since condition 3. is only asymptotic, but it is probably the simplest. The subspace $V(\undZ)$ in (A) comes from the
theorem in \cite{PT} that the above subspace with $Z_1=Z_2=Z_3=Z$ is asymptotically equal to the union of the lowest eigenspaces
of the full Schr\"odinger equation (\ref{SE}). In particular, this theorem dictates that the $V_i$ should consist
of Slater orbitals, not the commonly used Gaussian orbitals.
The presence of the variable dilation
parameters $Z_i$ and eq. (C) comes from requirement 2., which is equivalent to stationarity of the energy $\langle\Psi|H|\Psi\rangle$
of eigenstates
with respect to dilations (see the proof of Theorem \ref{T:Properties}).
%
%
%
%

(2) There is no empirical parameter.

(3) The model has the following variational formulation:
$$
   E^{CI} = \min_{\undZ\in(0,\infty)^3}\min_{\Psi\in V(\undZ), \, ||\Psi||=1} \langle\Psi | H | \Psi\rangle,
$$
with the set of minimizers $\Psi$ being equal to the set of normalized lowest eigenstates of (B). This is an immediate consequence
of (C) and the Rayleigh-Ritz variational principle for the bottom eigenvalue in (B).

(4) Dilation parameters like the $Z_i$ are closely related to physical ideas of screening, and go back at least to Slater
(in the context of the Hartree equations, \cite{Slater30, Slater64}). They are widely used in the quantum chemistry literature,
and are in most studies determined a priori, e.g. via a Hartree-Fock calculation (see \cite{Tayloretal1, SzaboOstlund96}).
However, from a mathematical standpoint
it is of interest to determine them variationally for each eigenstate, as done here; this implies that the ensuing wavefunctions satisfy
the virial theorem (see Theorem \ref{T:Properties}).
Note also that validity of the latter cannot be guaranteed by linear parameters (i.e., subspace enlargement), but requires making the model nonlinear.
This is because the dilation group $\Psi\mapsto Z^{3N/2}\Psi(Z\cdot)$, which underlies the virial theorem, is a non-compact group
which -- unlike the compact groups $SO(3)$ and $SU(2)$ corresponding to angular momentum and spin --
leaves {\it no} finite-dimensional subspace of $L^2_{anti}((\R^3\times\Z_2))$ invariant. The proof of this fact is left
to the interested reader.
\\[2mm]
We now extend (B), (C) to excited states. The simplest generalization would be to compute all eigenvalues $E_j(\undZ)$ and corresponding
orthonormal eigenstates $\Psi_j(\undZ)$ of $P(\undZ)HP(\undZ)$, then minimize each eigenvalue over $\undZ$.

But this procedure does not maintain the basic property of the full Hamiltonian (\ref{ham}) that eigenstates with different eigenvalue
are orthogonal. However, the symmetries described in Lemma \ref{L:FCIsymm} come to our help. If two eigenstates of the CI Hamiltonian
are also simultaneous eigenstates of the operators (\ref{ops}), which we can assume by Lemma \ref{L:FCIsymm}, then they
remain orthogonal after minimization of their eigenvalues over the $Z_i$, as long as the eigenvalue of at least
one of the operators (\ref{ops}) are different. Thus, in each symmetry subspace (i.e., each joint eigenstate of the operators (\ref{ops}))
we determine the values of the $Z_i$ that yield the minimum value for the lowest eigenvalue in the subspace, then use this
value to calculate all eigenvalues and eigenstates in the subspace. This way, orthogonality is maintained and in particular the CI
energy levels remain rigorous upper bounds to the true energy levels. In practice this method is very close to minimization of each eigenvalue,
since most symmetry subspaces turn out to be one-dimensional, and none are more than two-dimensional (see the next section).
The use of the $Z_i$'s from the lower state is of course a
somewhat arbitrary choice; it ensures the greatest accuracy possible for the lower lying states
(known as ``state-specific'' method), an alternative would be to choose the $Z_i$ so as to solve a least squares problem and
minimize the overall error.

To summarize, the {\bf minimal CI model for Li, Be, B, C, N, O, F, Ne excited states} is as follows. Below, $V^{L,S,P}(\undZ)$ denotes
the symmetry subspace $\{\Psi\in
V(\undZ)\, | \, \LL^2\Psi=L(L+1)\Psi, \, \SSS^2\Psi=S(S+1)\Psi, \, \hat{R}\Psi=p\Psi\}$, where $L$ is a non-negative integer, $S$ a non-negative
half-integer, and $p=\pm 1$.

\begin{table}[ht]
\resizebox{\textwidth}{!}{
  \begin{tabular}{|l l|}
   \hline
   (A') & (Choice of a parametrized, asymptotically exact family of subspaces) \\
        & As in (A) \\[2mm]
   (B') & (Subspace eigenvalue problem) For each symmetry subspace $V^{L,S,p}(\undZ)\subset V(\undZ)$ \\
       & $E^{CI}_j(\undZ) := \mbox{eigenvalues of }P(\undZ)HP(\undZ)\mbox{ on }V^{L,S,p}(\undZ)$, \\
       & $\Psi^{CI}_j(\undZ) := \mbox{corresponding orthonormal eigenstates}$, \\
       & $\mbox{where }P(\undZ)=\mbox{orthogonal projector of }L^2_a((\R^3\times\Z_2)^N)\mbox{ onto }V(\undZ)$ \\[2mm]
   (C') & (Variational parameter determination) For each symmetry subspace $V^{L,S,p}(\undZ)\subset V(\undZ)$ \\
       & $\undZ_* := \mbox{argmin}_{\undZ} (\min_j E^{CI}_j(\undZ)), \; E^{CI}_j := E^{CI}_j(\undZ_*), \; \Psi^{CI}_j := \Psi^{CI}_j(\undZ_*)$ \\[2mm]
   \hline
  \end{tabular}
} 
 \vspace*{-2mm}
 \label{Tab:MinimalCIExcited}
\end{table}

Let us summarize the additional properties of the model (A'), (B'), (C') beyond those of general symmetry-preserving CI (Lemmas \ref{L:UB}, \ref{L:FCIsymm})
in a theorem.
\begin{theorem} \label{T:Properties} Let $N\in\{3,..,10\}$, $Z\ge N$. The minimal CI model (A'), (B'), (C') has the following properties. \\[1mm]
(i) (Virial theorem) Any lowest normalized eigenstate $\Psi^{CI}$ of the model in a symmetry subspace (i.e., a
joint eigenspace of the symmetry operators $\LL^2$, $\SSS^2$, $\hat{R}$) satisfies
$$
     \langle\Psi^{CI}|V|\Psi^{CI}\rangle = -2\langle\Psi^{CI}|T|\Psi^{CI}\rangle,
$$
where $T$, $V$ are the kinetic respectively potential part of the Hamiltonian (\ref{ham}). \\[1mm]
(ii) (Correct asymptotic behaviour) For $N$ fixed and $Z\to\infty$,
$$
   \lim \frac{E^{CI}_j}{E_j} = 1, \;\;\; \lim \frac{\Delta E_j^{CI}}{\Delta E_j} = 1, \;\;\; \lim |||P^{CI}_j-P_j|||=0,
$$
where $E^{CI}_1<\dots<E^{CI}_m$ and $E_1<\dots<E_m$ are the CI eigenvalues respectively the lowest eigenvalues of the Schr\"odinger equation (\ref{SE}),
$\Delta E^{CI}$ and $\Delta E_j$ are the spectral gaps $E^{CI}_j-E^{CI}_1$ and $E_j-E_1$ ($j\ge 2$),
$P^{CI}_j$, $P_j$ denote the projectors onto the corresponding eigenspaces, and $|||\cdot|||$ is the operator norm on the
$N$-electron Hilbert space $L^2_a((\R^3\times\Z_2)^N)$.
\end{theorem}
\noindent
{\bf Proof} (i) follows from the fact that the manifold $\cup_{\undZ\in(0,\infty)^3}V^{L,S,p}(\undZ)$ is invariant under dilations
$\Psi\mapsto\Psi_\lambda(x_1,s_1,..,x_N,s_N)=\lambda^{3N/2}\Psi(\lambda x_1,s_1,..,
\lambda x_N,s_N)$, $\lambda>0$, which makes the usual proof of the virial theorem applicable: normalized minimizers
$\Psi$ of $\langle\Psi_\lambda|H|\Psi_\lambda\rangle$ in this manifold satisfy
$0=\frac{d}{d\lambda}\Big|_{\lambda=1}\langle\Psi_\lambda|H|\Psi_\lambda\rangle$.

(ii) is a consequence of the asymptotic results in \cite{PT} together with the elementary inequalities
$E_j\le E^{CI}_j\le E^{PT}_j$, where the $E^{PT}_j$ are the lowest eigenvalues of the PT model \cite{PT}.
\\[2mm]
We remark that statement (ii) fails when the Slater orbitals (\ref{CIorbitals}) are replaced by finite linear
combinations of Gaussians, or indeed by any functional form which fails to reproduce (\ref{CIorbitals}) asymptotically \cite{PT}.

It is instructive to compare the above argument in favour of Slater orbitals to the well known Kato cusp condition argument.
Theorem \ref{T:Properties} (ii) concerns the limit $N$ and $x$ general,
$Z\to\infty$, whereas the asymptotic regime of the Kato cusp condition
is $N$ and $Z$ general, $x\to 0$; the latter is therefore insufficient
to specify whole orbitals, as it only concerns their behaviour at $x=0$.
\\[2mm]
Finally, let us formulate a {\bf hierarchy of higher and higher dimensional CI models for the atom/ion with $N$ electrons}
which satisfy requirements 1., 2., 3. The models are parametrized by the number $n_{max}$ of
included single-electron ``shells'', and the only modification compared to (A'), (B'), (C') is an enlargement of
the family of subspaces $V(\undZ)$ in Step (A'), as follows.
For $n=1,...,n_{max}$, $\ell=0,..,n\!-\!1$, let $V_{n\ell}(\undZ) := \mbox{Span}\, \{R_{n\ell}(Z_1^\ell,..,Z_n^\ell,r)$ $\chi_{\ell
m}(\theta,\phi)\delta_\sigma(s)\, |$ $m=-\ell,..,\ell, \, \sigma=\pm\frac12\}$. Here $\undZ$ is the vector of dilation parameters
$Z_n^\ell$, $(r,\phi,\theta)$ are spherical polar coordinates in $\R^3$,
the $R_{n\ell}(Z_1^\ell,..,Z_n^\ell,\cdot)$ are orthonormal functions in $L^2(0,\infty)$ with respect to the measure $r^2dr$ which reduce
to the usual radial hydrogen eigenfunctions when $Z_1^\ell=...=Z_n^\ell=Z$, and the $\chi_{\ell m}$ are spherical harmonics (see \cite{PT}). Then take
$$
          V(\undZ):=\mbox{Span}\,\{\bigcup_d{\cal C}^d\},
$$
where $d=(d_{n\ell})_{n=1,..,n_{max}, \, \ell=0,..,n}$ runs over all partitions of $N$, i.e. $d_{n\ell}\ge 0$, $\sum_{n,\ell} d_{n\ell}=N$.
The minimal model (A'), (B'), (C') corresponds to taking $n_{max}=2$ (i.e., including only the first and second ``shell''), and imposing the
additional condition that the number $d_{1,0}$ of electrons in the subspace $V_{1,0}$ equals two (i.e., assuming that the first shell is
completely ``filled'').

\section{Minimal CI atomic energy levels and eigenstates} \label{Sec:Sol}

\subsection{Exact solution for given dilation parameters}
The key point allowing to solve the model (A'), (B'), (C') is the observation that the CI matrix $P(\undZ)HP(\undZ)$ in a simultaneous eigenbasis of $V(\undZ)$ of the symmetry operators (\ref{ops})
can be explicitly determined, and decouples into small invariant blocks. More precisely,
as noted in \cite{PT}, exact expressions can be derived for the joint eigenstates $\Psi_{\alpha}$ of (\ref{ops}) and their matrix elements $\langle\Psi_\alpha
| H | \Psi_\beta\rangle$
in terms of one-body, Coulomb and exchange integrals of the one-electron orbitals
(\ref{CIorbitals}); and when restricting without loss of generality to $S_3$ maximal and $L_3=0$ the
largest non-diagonal block is 2$\times$2.
For convenience we include the eigenfunctions and
symbolic matrix elements in Tables \ref{Tab:VeeMatrix1}--\ref{Tab:VeeMatrix2} below. The symmetry type of
the wavefunctions is also shown in Chemist's notation, which encodes the eigenvalues $L(L+1)$, $S(S+1)$ and $p$ of $\LL^2$, $\SSS^2$ and $\hat{R}$
by the symbol $^{2S+1}X^\nu$, where $L$ corresponds to $X$ via $0\to S$,. $1\to P$, $2\to D$, and no superscript $\nu$ means $p=1$, while
$\nu=o$ (for odd) stands for $p=-1$. Recall the standard notation for
one- and two-body integrals
\begin{equation}
     (a|b) = \langle a|hb\rangle, \;\; (ab | cd)  =  \int_{\R^6} dx_1 dx_2 a^*(x_1)b(x_1)\frac{1}{|x_1-x_2|} c^*(x_2)d(x_2),
    \label{ijklIntegral}
\end{equation}
where $h$ is the one-body Hamiltonian $-\tfrac{1}{2}\Delta - Z/|x|$.

\begin{table}[htbp]
 \begin{center}

  \resizebox{\textwidth}{!}{
   \begin{tabular}{|c|c|c|c|c|c|l|}
    \hline & $\LL^2$ &
    $\SSS^2$ &  $\hat{R}$ &   & $\Psi$
    & \multicolumn{1}{c|}{$\langle H \rangle$} \\
    \hline

    Li & $0$ & $\tfrac{3}{4}$ & $1$ & $^2S$ & $|1\overline{1}2\rangle$ & $2(1|1) + (2|2) + (11|11) + 2(11|22) - (12|21)$\\

    \cline{2-7} & $2$ & $\tfrac{3}{4}$ & $-1$ & $^2P^o$ & $|1\overline{1}3\rangle$ &
    $2(1|1) + (3|3) + (11|11) + 2(11|33)-(13|31)$
    \\
    \hline

    Be & $0$ & $0$ & $1$ & $^1S$ & $|1\overline{1}2\overline{2}\rangle$ &$2(1|1) + 2(2|2) + (11|11) +
    4(11|22) - 2(12|21) + (22|22)$\\

    \cline{6-7} & & & & &
    $\tfrac{1}{\sqrt{3}}\left(|1\overline{1}3\overline{3}\rangle +
    |1\overline{1}4\overline{4}\rangle +
    |1\overline{1}5\overline{5}\rangle \right)$ & $2(1|1) + 2(3|3) + (11|11) + 4(11|33)
    - 2(13|31) + (33|33) + 2(34|43)$ \\

    \cline{6-7} & & & & & cross &
     $\sqrt{3}(23|32)$ \\

    \cline{2-7} & $2$ & $0$ & $-1$ & $^1P^o$ &
    $\tfrac{1}{\sqrt{2}}\left(|1\overline{1}2\overline{3}\rangle
    -|1\overline{1}\overline{2}3\rangle\right)$ &  $2(1|1) + (2|2) + (3|3) + (11|11) + 2(11|22)
    - (12|21) +2(11|33)$ \\
    & & & & & & $- (13|31) + (22|33) + (23|32)$ \\

    \cline{2-7} & $2$ & $2$ & $-1$ & $^3P^o$ & $|1\overline{1}23\rangle$ & $2(1|1) + (2|2) + (3|3) + (11|11) +
    2(11|22) - (12|21) + 2(11|33)$ \\
    & & & & & & $- (13|31) + (22|33) - (23|32)$\\

    \cline{4-7} & & & $1$ & $^3P$ & $|1\overline{1}45\rangle$ &$2(1|1) + 2(3|3) + (11|11) + 4(11|33)
    -2(13|31) + (33|44) - (34|43)$ \\

    \cline{2-7} & $6$ & $0$ & $1$ & $^1D$ &
    $\tfrac{1}{\sqrt{6}}\left(2|1\overline{1}3\overline{3}\rangle-
    |1\overline{1}4\overline{4}\rangle
    -|1\overline{1}5\overline{5}\rangle \right)$
     &$2(1|1) + 2(3|3) + (11|11) + 4(11|33)
    -2(13|31) + (33|33) -(34|43)$\\

    \hline B & $0$ & $\tfrac{3}{4}$ & $1$ & $^2S$ &
    $\tfrac{1}{\sqrt{3}}\left(|1\overline{1}23\overline{3}\rangle +
    |1\overline{1}24\overline{4}\rangle +
    |1\overline{1}25\overline{5}\rangle\right)$&
     $2(1|1) + (2|2) + 2(3|3) + (11|11) + 2(11|22) - (12|21) + 4(11|33)$ \\
    & & & & & & $ -2(13|31) + 2(22|33) - (23|32) + (33|33) + 2(34|43)$\\

    \cline{2-7} & $0$ & $\tfrac{15}{4}$ & $-1$ & $^4S^o$ & $|1\overline{1}345\rangle$ &
    $2(1|1) + 3(3|3) + (11|11) + 6(11|33) -3(13|31) + 3(33|44) -3(34|43)$ \\

    \cline{2-7} & $2$ & $\tfrac{3}{4}$ & $-1$ & $^2P^o$ &
    $|1\overline{1}2\overline{2}{3}\rangle$ &
     $2(1|1) + 2(2|2) + (3|3) + (11|11) + 4(11|22) -2(12|21) + 2(11|33) $\\
    & & & & & & $ - (13|31) + (22|22) + 2(22|33) - (23|32)$ \\

    \cline{6-7} & & & & &
    $\tfrac{1}{\sqrt{2}}\left(|1\overline{1}34\overline{4}\rangle +
    |1\overline{1}35\overline{5}\rangle \right)$ & $ 2(1|1) + 3(3|3) + (11|11) +
    6(11|33) -3(13|31) + (33|33) + 2(33|44)$ \\

    \cline{6-7} & & & & & cross & $\sqrt{2}(23|32)$\\

    \cline{4-7} & & & $1$ & $^2P$ &
    $\tfrac{1}{\sqrt{6}}\left(2|1\overline{1}\overline{2}45\rangle
    -|1\overline{1}2\overline{4}5\rangle
    -|1\overline{1}24\overline{5}\rangle \right)$
    & $ 2(1|1) + (2|2) + 2(3|3) + (11|11) + 2(11|22) -(12|21) + 4(11|33)  $ \\
    & & & & & & $-2(13|31) + 2(22|33) + (23|32) + (33|44) -(34|43) $ \\

    \cline{2-7} & $2$ & $\tfrac{15}{4}$ & $1$ & $^4P$ & $|1\overline{1}245\rangle$ &
    $2(1|1) + (2|2) + 2(3|3) + (11|11) + 2(11|22) - (12|21) + 4(11|33) $ \\
    & & & & & & $-2(13|31) + 2(22|33) -2(23|32) + (33|44) - (34|43)$\\

    \cline{2-7} & $6$ & $\tfrac{3}{4}$ & $1$ & $^2D$ &
    $\tfrac{1}{\sqrt{6}}\left(2|1\overline{1}23\overline{3}\rangle
    -|1\overline{1}24\overline{4}\rangle
    -|1\overline{1}25\overline{5}\rangle \right)$
    & $2(1|1) + (2|2) + 2(3|3) + (11|11) + 2(11|22) -(12|21) + 4(11|33)$ \\
    & & & & & & $ -2(13|31) + 2(22|33) -(23|32) + (33|33) -(34|43)$ \\

    \cline{4-7} & & & $-1$ & $^2D^o$ &
    $\tfrac{1}{\sqrt{6}}\left(2|1\overline{1}\overline{3}45\rangle
    -|1\overline{1}3\overline{4}5\rangle
    -|1\overline{1}34\overline{5}\rangle \right)$ & $2(1|1) + 3(3|3) + (11|11) +
    6(11|33) -3(13|31) + 3(33|44)$ \\
    \hline

    C & $0$ & $0$ & $1$ & $^1S$ &
    $\tfrac{1}{\sqrt{3}}\left(|1\overline{1}2\overline{2}3\overline{3}\rangle
    + |1\overline{1}2\overline{2}4\overline{4}\rangle +
    |1\overline{1}2\overline{2}5\overline{5}\rangle \right) $
    & $2(1|1) + 2(2|2) + 2(3|3)0 + (11|11) + 4(11|22) -2(12|21) + 4(11|33) $ \\
    & & & & & & $-2(13|31) + (22|22) + 4(22|33) -2(23|32) + (33|33) + 2(34|43)$\\

    \cline{6-7} & & & & &
    $\tfrac{1}{\sqrt{3}}\left(|1\overline{1}3\overline{3}4\overline{4}\rangle
    + |1\overline{1}3\overline{3}5\overline{5}\rangle +
    |1\overline{1}4\overline{4}5\overline{5}\rangle \right) $
    & $2(1|1) + 4(3|3) + (11|11) + 8(11|33) -4(13|31) + 2(33|33) + 4(33|44)$ \\

    \cline{6-7} & & & & &
    cross
    &$2(23|32)$ \\

    \cline{2-7} & $0$ & $2$ & $-1$ & $^3S^o$ & $\hspace{-15mm} \tfrac{1}{\sqrt{12}}\big(
    3|1\overline{1}\overline{2}345\rangle -
    |1\overline{1}2\overline{3}45\rangle $
    & $2(1|1) + (2|2) + 3(3|3) + (11|11) + 2(11|22) - (12|21) + 6(11|33)  $ \\
    & & & & & $\hspace{15mm} -|1\overline{1}23\overline{4}5\rangle-
    |1\overline{1}234\overline{5}\rangle \big)$ &
    $-3(13|31) + 3(22|33) + (23|32) + 3(33|44) -3(34|43) $ \\

    \cline{2-7} & $0$ & $6$ & $-1$ & $^5S^o$ & $|1\overline{1}2345\rangle$ & $2(1|1) + (2|2) + 3(3|3) + (11|11) +
    2(11|22) - (12|21) + 6(11|33) $\\
    & & & & & & $-3(13|31) + 3(22|33) -3(23|32) + 3(33|44) -3(34|43)$ \\

    \cline{2-7} & $2$ & $0$ & $-1$ & $^1P^o$ &
    $\hspace{-15mm} \tfrac{1}{2}\big(|1\overline{1}2\overline{3}4\overline{4}\rangle -
    |1\overline{1}\overline{2}34\overline{4}\rangle$
    & $2(1|1) + (2|2) + 3(3|3) + (11|11) + 2(11|22) - (12|21) + 6(11|33)$ \\
    & & & & & $\hspace{15mm} +|1\overline{1}2\overline{3}5\overline{5}\rangle
    - |1\overline{1}\overline{2}35\overline{5}\rangle \big)$
    & $ -3(13|31) + 3(22|33) + (33|33) + 2(33|44)$ \\

    \cline{2-7} & $2$ & $2$ & $1$ & $^3P$ & $|1\overline{1}2\overline{2}45\rangle$
    &  $2(1|1) + 2(2|2)0 + 2(3|3) + (11|11) + 4(11|22) -2(12|21) + 4(11|33)$  \\
    & & & & & & $ -2(13|31) + (22|22) + 4(22|33) -2(23|32) + (33|44) - (34|43)$ \\

    \cline{6-7} & & & & & $|1\overline{1}3\overline{3}45\rangle$ &
    $2(1|1) + 4(3|3) + (11|11) + 8(11|33) -4(13|31) + (33|33) + 5(33|44)$ \\
    & & & & & & $-3(34|43)$\\

    \cline{6-7} & & & & & cross & $(23|32)$ \\

    \cline{4-7} & & & $-1$ & $^3P^o$ &
    $\tfrac{1}{\sqrt{2}}\left(|1\overline{1}234\overline{4}\rangle +
    |1\overline{1}235\overline{5}\rangle \right)$
    & $2(1|1) + (2|2) + 3(3|3) + (11|11) + 2(11|22) -(12|21) + 6(11|33)$ \\
    & & & & & & $ -3(13|31) + 3(22|33) -2(23|32) + (33|33) + 2(33|44)$ \\

    \cline{2-7} & $6$ & $0$ & $1$ & $^1D$ & $\tfrac{1}{\sqrt{6}} \left(
    2|1\overline{1}2\overline{2}3\overline{3}\rangle -
    |1\overline{1}2\overline{2}4\overline{4}\rangle -
    |1\overline{1}2\overline{2}5\overline{5}\rangle\right)$
    & $2(1|1) + 2(2|2) + 2(3|3) + (11|11) + 4(11|22) -2(12|21) + 4(11|33) $ \\
    & & & & & & $-2(13|31) + (22|22)+ 4(22|33) -2(23|32) + (33|33) -(34|43)$ \\

    \cline{6-7} & & & & & $\tfrac{1}{\sqrt{6}} \left(
    2|1\overline{1}4\overline{4}5\overline{5}\rangle -
    |1\overline{1}3\overline{3}4\overline{4}\rangle -
    |1\overline{1}3\overline{3}5\overline{5}\rangle \right)$ &
    $2(1|1) + 4(3|3) + (11|11) + 8(11|33) -4(13|31) + 2(33|33) + 4(33|44)$ \\
    & & & & & & $-3(34|43)$ \\

    \cline{6-7} & & & & & cross &
    $-(23|32)$\\

    \cline{4-7} & & & $-1$ & $^1D^o$ & $\tfrac{1}{\sqrt{12}} \big(
    2|1\overline{1}23\overline{4}\overline{5}\rangle
    -|1\overline{1}2\overline{3}4\overline{5}\rangle
    -|1\overline{1}2\overline{3}\overline{4}5\rangle$ &
    $2(1|1) + (2|2) + 3(3|3) + (11|11) + 2(11|22) - (12|21) + 6(11|33) $ \\
    & & & & & $+2|1\overline{1}\overline{2}\overline{3}45\rangle
    -|1\overline{1}\overline{2}34\overline{5}\rangle
    -|1\overline{1}\overline{2}3\overline{4}5\rangle \big)$
    & $-3(13|31) + 3(22|33) + 3(33|44)$\\

    \cline{2-7} & $6$ & $2$ & $-1$ & $^3D^o$ & $\tfrac{1}{\sqrt{6}} \big(
    2|1\overline{1}2\overline{3}45\rangle
    -|1\overline{1}234\overline{5}\rangle -
    |1\overline{1}23\overline{4}5\rangle\big)$
    & $2(1|1) + (2|2) + 3(3|3) + (11|11) + 2(11|22) - (12|21) + 6(11|33)$\\
    & & & & & &
    $ -3(13|31) + 3(22|33) -2(23|32) + 3(33|44)$\\
    \hline

   \end{tabular}

  } 

 \end{center}

 \caption
 {$H$ matrix element expressions for
 Li-C in subspace with $S_3$ maximal, $L_3=0$; `cross' denotes the
 off-diagonal term in the $2\times2$ matrix. Orbital notation is $1:=\phi_{1s}$, $2:=\phi_{2s}$, $3:=\phi_{2p_3}$, $4:=\phi_{2p_1}$,
$5:=\phi_{2p_2}$. Integrals $(a|a)$, $(ab|cd)$ are as in (\ref{ijklIntegral}).}
 \label{Tab:VeeMatrix1}
\end{table}

\begin{table}[htbp]
 \begin{center}

  \resizebox{\textwidth}{!}{
   \begin{tabular}{|c|c|c|c|c|c|l|}
     \hline & $\LL^2$&
    $\SSS^2$ &  $\hat{R}$&  & $\Psi$
    & \multicolumn{1}{c|}{$\langle H \rangle$} \\
    \hline

     N & $0$ & $\tfrac{3}{4}$ & $1$ & $^2S$ & $\tfrac{1}{\sqrt{3}} \big(
    |1\overline{1}23\overline{3}4\overline{4}\rangle +
    |1\overline{1}23\overline{3}5\overline{5}\rangle $
    & $2(1|1) + (2|2) + 4(3|3) + (11|11) + 2(11|22) - (12|21) + 8(11|33)$ \\
    & & & & & $+|1\overline{1}24\overline{4}5\overline{5}\rangle \big)$
    &  $ -4(13|31) + 4(22|33) -2(23|32) + 2(33|33) + 4(33|44)$ \\

    \cline{2-7} & $0$ & $\tfrac{15}{4}$ & $-1$ & $^4S^o$ &
    $|1\overline{1}2\overline{2}345\rangle$
     & $2(1|1) + 2(2|2) + 3(3|3) + (11|11) + 4(11|22) -2(12|21) + 6(11|33) $  \\
    & & & & & & $-3(13|31) + (22|22)+ 6(22|33) -3(23|32) + 3(33|44) -3(34|43)$ \\

    \cline{2-7} & $2$ & $\tfrac{3}{4}$ & $-1$ & $^2P^o$ &
    $\tfrac{1}{\sqrt{2}}\left(
    |1\overline{1}2\overline{2}34\overline{4}\rangle +
    |1\overline{1}2\overline{2}35\overline{5}\rangle \right)$  &
    $2(1|1) + 2(2|2) + 3(3|3) + (11|11) + 4(11|22) -2(12|21) + 6(11|33) $\\
    & & & & & & $-3(13|31) + (22|22)+ 6(22|33) -3(23|32) + (33|33) + 2(33|44)$ \\

    \cline{6-7} & & & & &
    $|1\overline{1}34\overline{4}5\overline{5}\rangle$ &
    $2(1|1) + 5(3|3) + (11|11) +
    10(11|33) -5(13|31) + 2(33|33) $ \\
    & & & & & & $+ 8(33|44) -4(34|43)$ \\

    \cline{6-7} & & & & & cross & $\sqrt{2}(23|32)$\\

    \cline{4-7} & & & $1$ & $^2P$ & $\tfrac{1}{\sqrt{6}} \big(
    2|1\overline{1}\overline{2}3\overline{3}45\rangle -
    |1\overline{1}23\overline{3}\overline{4}5\rangle $
    & $2(1|1) + (2|2) + 4(3|3) + (11|11) + 2(11|22) -(12|21) + 8(11|33)  $ \\
    & & & & & $- |1\overline{1}23\overline{3}4\overline{5}\rangle \big)$
    &  $-4(13|31) + 4(22|33) + (33|33) + 5(33|44) -3(34|43)$ \\

    \cline{2-7} & $2$ & $\tfrac{15}{4}$ & $1$ & $^4P$ &
    $|1\overline{1}23\overline{3}45\rangle$
    &  $2(1|1) + (2|2) + 4(3|3) + (11|11) + 2(11|22) - (12|21) + 8(11|33) $  \\
    & & & & & & $-4(13|31) + 4(22|33) -3(23|32) + (33|33) + 5(33|44) -3(34|43)$ \\

    \cline{2-7} & $6$ & $\tfrac{3}{4}$ & $-1$ & $^2D^o$ &
    $\tfrac{1}{\sqrt{6}}\big(2|1\overline{1}2\overline{2}\overline{3}45\rangle
    - |1\overline{1}2\overline{2}34\overline{5}\rangle$
    & $2(1|1) + 2(2|2) + 3(3|3) + (11|11) + 4(11|22) -2(12|21) + 6(11|33)$  \\
    & & & & & $- |1\overline{1}2\overline{2}3\overline{4}5\rangle \big)$
    & $ -3(13|31) + (22|22)+ 6(22|33) -3(23|32) + 3(33|44)$ \\

    \cline{4-7} & & & $1$ & $^2D$ &
    $\tfrac{1}{\sqrt{6}}\big(2|1\overline{1}24\overline{4}5\overline{5}\rangle
    - |1\overline{1}23\overline{3}4\overline{4}\rangle$
    & $2(1|1) + (2|2) + 4(3|3) + (11|11) + 2(11|22) -(12|21) + 8(11|33)$ \\
    & & & & & $-|1\overline{1}23\overline{3}5\overline{5}\rangle\big)$
    & $ -4(13|31) + 4(22|33) -2(23|32) + 2(33|33) + 4(33|44) -3(34|43)$ \\

    \hline

    O & $0$ & $0$ & $1$ &  $^1S$ & $\tfrac{1}{\sqrt{3}}\big(
    |1\overline{1}2\overline{2}3\overline{3}4\overline{4}\rangle +
    |1\overline{1}2\overline{2}3\overline{3}5\overline{5}\rangle$ &
    $2(1|1) + 2(2|2) + 4(3|3) + (11|11) +
    4(11|22) -2(12|21) + 8(11|33)$\\
    & & & & & $+
    |1\overline{1}2\overline{2}4\overline{4}5\overline{5}\rangle\big)$
    & $ -4(13|31) +
    (22|22) + 8(22|33) -4(23|32) + 2(33|33) + 4(33|44)$\\

    \cline{6-7} & & & & & $|1\overline{1}3\overline{3}4\overline{4}5\overline{5}\rangle$ &
    $2(1|1) + 5(3|3) + (11|11) + 12(11|33) -6(13|31) + 3(33|33))$\\
    & & & & & & $ + 12(33|44 -6(34|43)$ \\

    \cline{6-7} & & & & & cross &
     $\sqrt{3}(23|32)$ \\

    \cline{2-7} & $2$ & $0$ & $-1$ & $^1P^o$ & $\tfrac{1}{\sqrt{2}}\left(
    |1\overline{1}2\overline{3}4\overline{4}5\overline{5}\rangle-
    |1\overline{1}\overline{2}34\overline{4}5\overline{5}\rangle\right)$
    & $2(1|1) + (2|2) + 5(3|3) + (11|11) + 2(11|22) -(12|21) + 10(11|33) $ \\
    & & & & & & $-5(13|31) + 5(22|33) -(23|32) + 2(33|33) + 8(33|44) -4(34|43)$ \\

    \cline{2-7} & $2$ & $2$ & $-1$ & $^3P^o$ &
    $|1\overline{1}234\overline{4}5\overline{5}\rangle$
    & $2(1|1) + (2|2) + 5(3|3) + (11|11) + 2(11|22) - (12|21) + 10(11|33) $ \\
    & & & & & & $-5(13|31) + 5(22|33)-3(23|32) + 2(33|33) + 8(33|44) -4(34|43)$ \\

    \cline{4-7} & & & $1$ & $^3P$ &
    $|1\overline{1}2\overline{2}3\overline{3}45\rangle$
    & $2(1|1) + 2(2|2) + 4(3|3) + (11|11) + 4(11|22) -2(12|21) + 8(11|33) $ \\
    & & & & & & $-4(13|31) + (22|22) + 8(22|33) -4(23|32) + (33|33) + 5(33|44)$ \\
    & & & & & & $ -3(34|43)$ \\

    \cline{2-7} & $6$ & $0$ & $1$ & $^1D$ & $\tfrac{1}{\sqrt{6}}\big(
    2|1\overline{1}2\overline{2}4\overline{4}5\overline{5}\rangle-
    |1\overline{1}2\overline{2}3\overline{3}4\overline{4}\rangle$
    & $2(1|1) + 2(2|2) + 4(3|3) + (11|11) + 4(11|22) -2(12|21) + 8(11|33) $ \\
    & & & & &
    $-|1\overline{1}2\overline{2}3\overline{3}5\overline{5}\rangle
    \big)$
    & $-4(13|31) + (22|22) + 8(22|33) -4(23|32) + 2(33|33) + 4(33|44)$\\
    & & & & & & $-3(34|43)$\\
    \hline

    F & $0$ & $\tfrac{3}{4}$ & $1$ & $^2S$ &
    $|1\overline{1}23\overline{3}4\overline{4}5\overline{5}\rangle$
    & $2(1|1) + (2|2) + 6(3|3) + (11|11) + 2(11|22) - (12|21) + 12(11|33) $  \\
    & & & & & & $-6(13|31) + 6(22|33) -3(23|32) + 3(33|33) + 12(33|44) -6(34|43)$ \\

    \cline{2-7} & $2$ & $\tfrac{3}{4}$ & $-1$ & $^2P^o$ &
    $|1\overline{1}2\overline{2}34\overline{4}5\overline{5}\rangle$
    & $2(1|1) + 2(2|2) + 5(3|3) + (11|11) + 4(11|22) -2(12|21) + 10(11|33) $  \\
    & & & & & & $-5(13|31) + (22|22) + 10(22|33) -5(23|32) + 2(33|33) + 8(33|44)$\\
    & & & & & & $ -4(34|43)$ \\
    \hline

    Ne & $0$ & $0$ & $1$ & $^1S$ &
    $|1\overline{1}2\overline{2}3\overline{3}4\overline{4}5\overline{5}\rangle$
    & $2(1|1) + 2(2|2) + 6(3|3) + (11|11) + 4(11|22) -2(12|21)+12(11|33)$\\
    & & & & & & $-6(13|31) +(22|22)+12(22|33)-6(23|32)+3(33|33) + 12(33|44)$ \\
    & & & & & & $-6(34|43)$ \\
    \hline

   \end{tabular}
  }  

 \end{center}

 \caption
 {$H$ matrix element expressions for
 N-Ne in subspace with $S_3$ maximal, $L_3=0$; `cross' denotes the
 off-diagonal term in the $2\times2$ matrix. Orbital notation is $1:=\phi_{1s}$, $2:=\phi_{2s}$,
 $3:=\phi_{2p_3}$, $4:=\phi_{2p_1}$, $5:=\phi_{2p_2}$. Integrals $(a|a)$, $(ab|cd)$ are as in (\ref{ijklIntegral}).}
 \label{Tab:VeeMatrix2}
\end{table}

It remains to evaluate the one-body Coulomb and exchange integrals for the basis (\ref{CIorbitals}).
Despite the basis not being Gaussian, they can be evaluated exactly, by
the method introduced in \cite{PT}: by Fourier calculus, we can re-write
$(ab|cd) = (2\pi^2)^{-1}\int_{\R^3}|k|^{-2}(\widehat{ab^*})^*\widehat{c^*d}$;
we then derive the Fourier transform of the pointwise products of the orbitals (\ref{CIorbitals}) (see Lemma \ref{L:FT}),
reduce to 1D integrals with the help of spherical polar coordinates in $k$-space, and evaluate the
remaining 1D integrals -- whose integrands turn out to be rational functions -- via the residue theorem
(or MAPLE). The result is as follows.
\begin{lemma} \label{L:FT}
The Fourier transforms of pointwise products of the one-electron orbitals
(\ref{CIorbitals}) are as follows.  In
all cases $j, \ell =1,2,3$, $j \neq \ell$.
 \begin{center}
 \resizebox{0.7\textwidth}{!}{
  \begin{tabular}{|c|c|}
    \hline Function & Fourier Transform \\
    \hline $\phi_{1s}\phi_{1s}$ & $\frac{16Z_1^4}{(4Z_1^2+|k|^2)^2}$ \\

    $\phi_{2s}\phi_{2s}$ &
    $\frac{Z_2^5}{(4Z_1^2-2Z_1Z_2+Z_2^2)}\left(\frac{2(Z_1+2Z_2)}{(Z_2^2+|k|^2)^2}-
    \frac{Z_2(2Z_1+Z_2)(2Z_1+5Z_2)}{(Z_2^2+|k|^2)^3} +
    \frac{2Z_2^3(2Z_1+Z_2)^2}{(Z_2^2+|k|^2)^4}\right)$\\

    $\phi_{1s}\phi_{2s}$ &
    $\frac{\sqrt{6}Z_1^{3/2}Z_2^{5/2}}{\sqrt{(4Z_1^2-2Z_1Z_2+Z_2^2)}}
    \left(\frac{4(2Z_1+Z_2)}{3((Z_1+Z_2/2)^2+|k|^2)^2}
    -\frac{(2Z_1+Z_2)^3}{3((Z_1+Z_2/2)^2+|k|^2)^3}\right)$ \\

    $\phi_{2p_j}\phi_{2p_j}$& $\frac{Z_3^6}{(Z_3^2+|k|^2)^3}-\frac{6Z_3^6k_j^2}{(Z_3^2+|k|^2)^4}$ \\

    $\phi_{1s}\phi_{2p_j}$& $-
    \frac{2\sqrt{2}iZ_1^{3/2}Z_3^{5/2}
    (2Z_1+Z_3)k_j}{\left((Z_1+Z_3/2)^2+|k|^2\right)^3}$ \\

    $\phi_{2s}\phi_{2p_j}$& $\frac{\sqrt{3}iZ_2^{5/2}Z_3^{5/2}}
    {16\sqrt{4Z_1^2-2Z_1Z_2+Z_2^2}} \left( \frac{8(Z_2+Z_3)^2(2Z_1+Z_2)k_j}
    {\left( ((Z_2+Z_3)/2)^2+|k|^2\right)^4} - \frac{(32Z_1+64Z_2+48Z_3)k_j}
    {3\left( ((Z_2+Z_3)/2)^2+|k|^2\right)^3} \right)$ \\

    $\phi_{2p_j}\phi_{2p_\ell}$& $-\frac{6k_jk_{\ell}Z_3^6}{(Z_3^2+|k|^2)^4}$ \\
    \hline
  \end{tabular}
 } 
 \end{center}
\end{lemma}

\begin{lemma}\label{L:VeeElements}
Using the abbreviated notation
$1=\phi_{1s}$, $2=\phi_{2s}$, $3=\phi_{2p_3}$, $4=\phi_{2p_1}$, the one-body, Coulomb and
exchange integrals (\ref{ijklIntegral}) for the orbitals (\ref{CIorbitals}) respectively the PT orbitals ($Z_1=Z_2=Z_3=Z$) are given
by
 \begin{center}
 \resizebox{0.9\textwidth}{!}{
  \begin{tabular}{|c|c|c|}
    \hline
    Integral & CI value & PT value \\
    \hline
     $(1|1)$  &    $\frac{1}{2}Z_1^2 -ZZ_1$            &  $-\frac12 Z^2$     \\
     $(2|2)$  &    $\frac{Z_2^2}{24}\frac{4Z_1^2-2Z_1Z_2+7Z_2^2}
    {4Z_1^2-2Z_1Z_2+Z_2^2} -\frac{ZZ_2}{4}\frac{4Z_1^2-4Z_1Z_2+3Z_2^2}{4Z_1^2-2Z_1Z_2+Z_2^2}$    &     $-\frac18 Z^2$    \\
     $(3|3)$  &    $\frac{1}{8}Z_3^2-\frac{1}{4}ZZ_3$  &  $-\frac18 Z^2$     \\
    \hline
    $(11|11)$ & $\frac{5}{8}Z_1$ & $\frac{5}{8}Z$ \\

    $(11|22)$ & $\frac{Z_1Z_2(8Z_1^4+4Z_1^3Z_2+4Z_1Z_2^3+Z_2^4)}{(2Z_1+Z_2)^3(4Z_1^2-2Z_1Z_2+Z_2^2)}$ &
                                   $\frac{17}{81}Z$ \\

    $(12|21)$ & $\frac{16Z_1^3Z_2^5}{(4Z_1^2-2Z_1Z_2+Z_2^2)(2Z_1+Z_2)^5}$ &
                                   $\frac{16}{729}Z$ \\

    $(22|22)$ & $\frac{Z_2}{512}\frac{(1488 Z_1^4-1952 Z_1^3 Z_2 + 1752 Z_1^2 Z_2^2 -840Z_1Z_2^2 + 245Z_2^4)}
    {(4Z_1^2-2Z_1Z_2+Z_2^2)^2}$ &
                                   $\frac{77}{512}Z$ \\

    $(11|33)$ & $\frac{Z_1Z_3\left( 8Z_1^4 + 20Z_1^3Z_3 + 20 Z_1^2Z_3^2
    + 10Z_1Z_3^3 +Z_3^4\right)}{(2Z_1+Z_3)^5}$ &
                                   $\frac{59}{243}Z$ \\

    $(13|31)$ & $\frac{112 Z_1^3 Z_3^5}{3 (2Z_1+Z_3)^7}$ &
                                   $\frac{112}{6561}Z$ \\

    $(22|33)$ & $\frac{Z_2Z_3}{(4(4Z_1^2-2Z_1Z_2+Z_2^2)(Z_2+Z_3)^7)}$
    {\scriptsize $ \bigl[(4Z_1^2-2Z_1Z_2+Z_2^2)(Z_2^6+7Z_2^5Z_3+21Z_2^4
    Z_3^2+35Z_2^3Z_3^3)$ }
    & \\
    & {\scriptsize $+ 3Z_2^2Z_3^4(28Z_1^2-28Z_1Z_2+11Z_2^2)$} {\scriptsize $+ 7Z_2Z_3^5(4Z_1^2-4Z_1Z_2+3Z_2^2)+Z_3^6(4Z_1^2-4Z_1Z_2+3Z_2^2)
    ) \bigr]$ } &
                                    $\frac{83}{512}Z$ \\

    $(23|32)$ &
    $\frac{Z_2^5Z_3^5(740Z_1^2+152Z_1Z_2+17Z_2^2-42Z_2Z_3-588Z_1Z_3+
    126Z_3^2)}{(9(Z_2+Z_3)^9(4Z_1^2-2Z_1Z_2+Z_2^2))}$ &
                                    $\frac{15}{512}Z$ \\

    $(33|33)$ & $\frac{501}{2560}Z_3$ &
                                    $\frac{501}{2560} Z$ \\

    $(33|44)$ & $\frac{447}{2560}Z_3$ &
                                    $\frac{447}{2560} Z$ \\

    $(34|43)$ & $\frac{27}{2560}Z_3$ &
                                    $\frac{27}{2560} Z$ \\
    \hline
  \end{tabular}
 } 
 \end{center}
\end{lemma}

This table together with Tables \ref{Tab:VeeMatrix1}--\ref{Tab:VeeMatrix2} yields, for any given values of the $Z_i$,
the exact solution of the linear part (B') of the CI model in the nondegenerate symmetry subspaces.

In the 2D subspaces, the tables need to be combined with the analytic expression for the eigenvalues
of the $2\times 2$ matrices $(\langle\Psi_i|H|\Psi_j\rangle)_{i,j=1}^2$
(see \cite{PT} and denote $\langle\Psi_i|H|\Psi_j\rangle=:H_{ij}$),
\be \label{degevals}
    \lambda_{\pm} = \frac{H_{11}+H_{22}}{2} \pm
    \sqrt{\big(\tfrac{H_{11}-H_{22}}{2}\big)^2 +
    |H_{12}|^2},
\ee
and corresponding normalized eigenstates,
\be \label{degestates}
    \Psi_\pm = \frac{1}{\sqrt{1+c_\pm^2}}\Bigl( \Psi_1 + c_\pm \Psi_2 \Bigr), \;\;\;
    c_\pm = \frac{ \frac{H_{22}-H_{11}}{2} \pm  \sqrt{\big(\tfrac{H_{22}-H_{11}}{2}\big)^2 +
    |H_{12}|^2     }          }{H_{12}}.
\ee
Thus we have analytic expressions for all eigenvalues and eigenvectors of $P(\undZ)HP(\undZ)$ in terms of the $Z_i$.
\subsection{Numerical optimization of dilation parameters}
The final stage is to minimize the exact energy levels over the $Z_i$ (Step (C') of the minimal CI model), which
is performed using MAPLE. Since we are dealing with only a 3-parameter minimization over explicit rational or square root functions,
we obtain highly accurate numerical energy
levels, along with their eigenspaces and symmetries.
In particular, all digits indicated in the Tables below are believed to be exact relative to the underlying model (A'), (B'), (C').

\subsection{Final result}

The minimal CI energy levels, along with the minimizing values of the dilation parameters
$Z_i$, for $N=3, \dots, 10$, are shown in Tables
\ref{Tab:EnergyComparison1} and \ref{Tab:EnergyComparison2}. The corresponding eigenspaces are as given in Tables \ref{Tab:VeeMatrix1}--\ref{Tab:VeeMatrix2}.
%

\begin{table}[h!tbp]
 \begin{center}
\resizebox{!}{0.8\textheight}{
\rotatebox{90}{
  \begin{tabular}{|c|c|c|c|c|c|c|c|c|c|c|c|c|}
    \hline & State & \multicolumn{1}{|c|}{$E_{CI}$}
    & \multicolumn{1}{|c|}{$Z_1$} & \multicolumn{1}{|c|}{$Z_2$}
    & \multicolumn{1}{|c|}{$Z_3$} & \multicolumn{1}{|c|}{$c$}
    & \multicolumn{1}{|c|}{$E_{Exp}$} & \multicolumn{1}{|c|}{$E_{MDHF}$} &
    \multicolumn{1}{|c|}{$E_{PT}$} &
    \multicolumn{1}{|c|}{$\Delta E_{CI}$} &
    \multicolumn{1}{|c|}{$\Delta E_{Exp}$} &
    \multicolumn{1}{|c|}{$\Delta E_{MDHF}$}\\
    \hline
    Li & $^2S$ & -7.4139 & 2.6937 &1.5334 & & & -7.4779 & -7.4327 &-7.0566 & & & \\
    & $^2P^\circ$ & -7.3504 &  2.6858 & & 1.0458  &
    & -7.4100 & -7.3651 & -6.8444 & 0.0635 & 0.0679 & 0.0677 \\
    \hline
    Be & $^1S$ & -14.5795 & 3.7052 & 2.3669 & 1.9944 & -0.3597 & -14.6684 & -14.5730 &-13.7629 &  &  &\\
    & $^3P^\circ$ & -14.4823 & 3.6944 & 2.4045 & 1.7807 & & -14.5683 & -14.5115 &-13.5034 & 0.0972 & 0.1001 & 0.0615\\
    & $^1P^\circ$ & -14.3688 & 3.6962 & 2.6684 & 0.9324 & & -14.4745 & -14.3947 &-13.2690 & 0.2107 & 0.1939 & 0.1783\\
    & $^1D$ & -14.2764 & 3.6813 & & 1.7025 & & -14.4092 &          & -13.0112 & 0.3030 & 0.2592 &       \\
    & $^3P$ & -14.3128 & 3.6806 & & 1.7502 & & -14.3964 &          & -13.0955 & 0.2667 & 0.2720 &       \\
    & $^1S$ & -14.1439 & 3.7052 & 2.3669 & 1.9944 & 2.7802 & (-14.3212) &          &-12.8377 &  0.4356 & (0.3471) &       \\
    \hline
    B & $^2P^\circ$ & -24.4885 & 4.7086 & 3.1628 & 2.4660 & -0.2664 & -24.6581 & -24.5291 & -22.7374 & & &\\
    & $^4P$ & -24.3969 & 4.6925 & 3.2440 & 2.4757 & & -24.5265 & -24.4507 & -22.4273 & 0.0915 & 0.1316 & 0.0784\\
    & $^2D$ & -24.2448 & 4.6930 & 3.2432 & 2.3470 & & -24.4401 & -24.3119 & -22.1753 & 0.2437 & 0.2181 & 0.2172\\
    & $^2S$ & -24.1719 & 4.6938 & 3.2710 & 2.2573 & & (-24.3685) & -24.2481 & -22.0171 & 0.3165 & (0.2896) & 0.2810\\
    & $^2P$ & -24.1010 & 4.6932 & 3.3746 & 2.1187 & & -24.3276 & -24.1790 & -21.9878 & 0.3875 & 0.3305 & 0.3500\\
    & $^4S^\circ$ & -24.0776 & 4.6732 &        & 2.4432 & & -24.2157 &          & -21.7612 & 0.4807 & 0.4424 & \\
    & $^2D^\circ$ & -24.0010 & 4.6742 &        & 2.3960 & & (-24.2034) &          &-21.6030 &  0.4876 & (0.4547) & \\
    & $^2P^\circ$ & -23.9076 & 4.7086 & 3.1628 & 2.4660 & 3.7536  & (-24.1319) &          &-21.4629 & 0.5808 & (0.5062) & \\
    \hline
    C & $^3P$ & -37.5689 & 5.7107 & 3.9670 & 3.1116 & -0.1706  & -37.8558 & -37.6886 & -34.4468 & & &  \\
    & $^1D$ & -37.5039 & 5.7114 & 3.9790 & 3.0520 & 0.1690  & -37.8094 & -37.6313 &-34.3202 & 0.0650 & 0.0464 & 0.0573\\
    & $^1S$ & -37.4656 & 5.7096 & 3.9998 & 3.0265 & -0.3126 & -37.7572 & -37.5496 &34.1838 & 0.1033 & 0.0986 & 0.1390\\
    & $^5S^\circ$ & -37.4974 & 5.6893 & 4.0713 & 3.1623 & & -37.7021 & -37.5992 & -34.0859 & 0.0715 & 0.1537 & 0.0894\\
    & $^3D^\circ$ & -37.2698 & 5.6894 & 4.0501 & 3.0739 & & -37.5638 & -37.3944 &-33.7203 & 0.2991 & 0.2920 & 0.2945\\
    & $^3P^\circ$ & -37.2053 & 5.6899 & 4.0599 & 3.0389 & & (-37.5129) & -37.3377 &-33.5938 & 0.3636 & (0.3429) & 0.3509\\
    & $^1D^\circ$ & -37.0173 & 5.6885 & 4.0265 & 2.9773 & & (-37.4100) & -37.1696 &-33.3688 & 0.5516 & (0.4458) & 0.5190\\
    & $^3S^\circ$ & -36.9869 & 5.6873 & 3.9731 & 2.9938 & & -37.3737 & -37.1421 &-33.3828 & 0.5820 & 0.4821 & 0.5465\\
    & $^1P^\circ$ & -36.9550 & 5.6892 & 4.0577 & 2.9316 & & (-37.3096) & -37.1158 &-33.2422 & 0.6139 & (0.5462) & 0.5728\\
    & $^3P$ & -36.7965 & 5.7107 & 3.9670 & 3.1116 & 5.8631 &          &          &-32.7641 & 0.7724 &        &       \\
    & $^1D$ & -36.7331 & 5.7114 & 3.9790 & 3.0520 & -5.9172 &          &          &-32.6376 & 0.8358 &        &       \\
    & $^1S$ & -36.5799 & 5.7096 & 3.9998 & 3.0265 & 3.1994 &          &          &-32.3943 & 0.9889 &        &       \\
    \hline
  \end{tabular}
}
}
 \end{center}
 \caption{Energy levels for minimal asymptotics-based CI, rounded to 4 decimal places (a.u.).
 $Z_i$ are the minimizing dilation parameters and $c$ is the correlation coefficient
 in the eigenfunctions.  Also shown are Multi-Determinant Hartree-Fock
 \protect\cite{Tatewaki94} and experimental energies \cite{NIST}. $\Delta E$ is
 the energy difference to respective ground state . Parentheses
denote levels assigned to the appropriate configuration (by comparison with
numerical calculations) but which lie above levels of the same symmetry not
shown in the table.(Li--C).}
 \label{Tab:EnergyComparison1}
\end{table}

\begin{table}[h!tbp]
 \begin{center}
\resizebox{!}{0.8\textheight}{
\rotatebox{90}{
  \begin{tabular}{|c|c|c|c|c|c|c|c|c|c|c|c|c|}
    \hline & State & \multicolumn{1}{|c|}{$E_{CI}$}
    & \multicolumn{1}{|c|}{$Z_1$} & \multicolumn{1}{|c|}{$Z_2$}
    & \multicolumn{1}{|c|}{$Z_3$} & \multicolumn{1}{|c|}{$c$}
    & \multicolumn{1}{|c|}{$E_{Exp}$} & \multicolumn{1}{|c|}{$E_{MDHF}$} &
    \multicolumn{1}{|c|}{$E_{PT}$} &
    \multicolumn{1}{|c|}{$\Delta E_{CI}$} &
    \multicolumn{1}{|c|}{$\Delta E_{Exp}$} &
    \multicolumn{1}{|c|}{$\Delta E_{MDHF}$}\\
    \hline
    N & $^4S^\circ$ & -54.1597 & 6.7117 & 4.7535 & 3.7924 & & -54.6117 & -54.4009 &-49.1503 & & & \\
    & $^2D^\circ$ & -54.0407 & 6.7124 & 4.7711 & 3.7317 & & -54.5241 & -54.2962 &-48.9288 & 0.1190 & 0.0876 & 0.1048\\
    & $^2P^\circ$ & -54.0075 & 6.7110 & 4.7893 & 3.7162 & -0.2091  & -54.4803 & -54.2281 &-48.8195 & 0.1523 & 0.1314 & 0.1728\\
    & $^4P$ & -53.7666 & 6.6854 & 4.8658 & 3.7592 &  & (-54.2101) & -53.9883 & -48.1630 & 0.3932 & (0.4016) & 0.4127\\
    & $^2D$ & -53.5340 & 6.6850 & 4.8414 & 3.7065 & & (-54.0595) & -53.7836 & -47.8103 & 0.6257 & (0.5522) & 0.6173\\
    & $^2S$ & -53.4173 & 6.6857 & 4.8575 & 3.6669 & &          & -53.6834 & -47.5888 & 0.7424 &        & 0.7175\\
    & $^2P$ & -53.3071 & 6.6830 & 4.7591 & 3.6794 & &          & -53.5839 & -47.5478 & 0.8526 &        & 0.8170\\
    & $^2P^\circ$ & -52.9277 & 6.7110 & 4.7893 & 3.7162 & 4.7815  &          &          &-46.5905 & 1.2320 &        &       \\
    \hline
    O & $^3P$ & -74.3931 & 7.7118 & 5.5613 & 4.4117 &  & -75.1080 & -74.8094 &-66.7048 & & &\\
    & $^1D$ & -74.3004 & 7.7122 & 5.5709 & 4.3828 & & -75.0357 & -74.7293 & -66.5360 & 0.0928 & 0.0723 & 0.0801\\
    & $^1S$ & -74.2328 & 7.7103 & 5.5967 & 4.3628 & -0.2283  & -74.9540 & -74.6110 & -66.3421 & 0.1603 & 0.1540 & 0.1984\\
    & $^3P^\circ$ & -73.7784 & 7.6805 & 5.6490 & 4.3916 & & (-74.5324) & -74.1839 &-65.3265 & 0.6147 & (0.5756) & 0.6255\\
    & $^1P^\circ$ & -73.4204 & 7.6785 & 5.5620 & 4.3549 & &          & -73.8720 &-64.8578 & 0.9727 &        & 0.9374\\
    & $^1S$ & -72.8054 & 7.7103 & 5.5967 & 4.3628 & 4.3811  &          &          &-63.4984 & 1.5877 &        &       \\
    \hline
    F & $^2P^\circ$ & -98.7503 & 8.7112 & 6.3576 & 5.0587 & & -99.8060 & -99.4093 & -87.6660& & &  \\
    & $^2S$ & -97.8704 & 8.6748 & 6.4189 & 5.0416 & & (-99.0322) & -98.5312 &-85.8342& 0.8800 & (0.7738) & 0.8781\\
    \hline
    Ne & $^1S$ & -127.5695 & 9.7101 & 7.1469 & 5.7177 & & -129.0500 & -128.5471 &-112.2917& & &  \\
    \hline
   \end{tabular}
}
}
 \end{center}
 \caption{Energy levels for minimal asymptotics-based CI, rounded to 4 decimal places (a.u.).
 $Z_i$ are the minimizing dilation parameters and $c$ is the correlation coefficient
 in the eigenfunctions.  Also shown are Multi-Determinant Hartree-Fock
 \protect\cite{Tatewaki94} and experimental energies \cite{NIST}. $\Delta E$ is
 the energy difference to the respective ground state. Parentheses
denote levels assigned to the appropriate configuration (by comparison with
numerical calculations) but which lie above levels of the same symmetry not
shown in the table. (N--Ne).}
 \label{Tab:EnergyComparison2}
\end{table}

\section{Comparison with large-scale numerical calculations and experiment} \label{Sec:Comparison}
\subsection{Ground state energies and ground states} \label{Sec:GSE}

The results in Tables \ref{Tab:EnergyComparison1} and \ref{Tab:EnergyComparison2} show that the symmetry of the ground state
of the model (A), (B), (C) agrees with experiment in every case,
and that the ground state energies capture around 99~$\%$ of the experimental energy.

We consider this agreement very good for such a low-dimensional projection of the Schr\"odinger equation. In the case of Beryllium,
our ground state CI energy even outperforms the benchmark numerical multi-determinant Hartree-Fock results of \cite{Tatewaki94}.
This demonstrates that a careful choice of basis and
considering the full Hamiltonian, including all correlation terms, can be more effective than large numerical computations.

It is also of theoretical interest to compare with the best numerical values in the literature, which rely on
more high-powered approaches. The table below compares, in a typical example,
our asymptotics-based minimal CI results, the MDHF results of Tatewaki et al. (also based on a small number of
determinants but on a huge one-electron basis set, considered essentially complete), the MPII results of
Canal Neto, Muniz, Centoducatte and Jorge, and the benchmark Full CI results
of Bauschlicher and Taylor.

\begin{table}[http]
\resizebox{\textwidth}{!}{
  \begin{tabular}{|l|c|c|c|c|c|}
   \hline
   Method & 1st order PT \cite{PT} & minimal CI (this paper) & MDHF \cite{Tatewaki94}  & MPII \cite{CanalNeto05} & FCI \cite{Tayloretal1}
\\
   \hline
   DOF's  & 8  & 11 & $\sim 10^3$ (estimate) & not given & 2.8$\times$10$^7$
\\
   \hline
   Error  & 12$\%$ & 1.06$\%$ & 0.40$\%$ & 0.28$\%$ & 0.21$\%$
\\
   \hline
  \end{tabular}
} 
\vspace*{-2mm}
\caption{Fluorine atom ground state energy (various methods). DOF's means the number of computational
degrees of freedom.}
 \label{Tab:variousmethods}
\end{table}

Other examples we considered gave a similar picture. In particular, for $N\ge 5$ asymptotics-based minimal CI was never outperformed by more than
one digit in all tested cases, not even by the recent explicitly correlated, multi-configurational variational Monte Carlo results \cite{Galvez02};
for $N=4$ (Be) the sophisticated iterative subspace recursions of \cite{Bendazzoli04, Nakatsuji07}
 -- which lead to
complicated final wavefunctions with $3\times 10^6$ respectively $2\times 10^3$ DOF's --  only yield energies which are one respectively two digits more accurate.
While from an applications point of view
an accuracy gain of one digit can be very important, the fact remains that the required computational effort is larger by
many orders of magnitude. A tentative conclusion is that a significant part
of the quality of quantum chemistry models lies in making a sophisticated initial ansatz, while subsequent efforts
to include more and more contributions appear to
exhibit the same disappointing scaling behaviour expected from a direct discretization of a problem suffering from
the curse of dimension.
\\[2mm]
Also of theoretical interest is the large gain in accuracy of minimal CI over the PT model (i.e. first order
perturbation theory with respect to electron interaction) \cite{PT},
since the two models differ only by the optimization step (C) over dilation parameters.

\begin{table}[htbp]
 \begin{center}
  \resizebox{0.8\textwidth}{!}{
  \begin{tabular}{|l|c|c|c|c|c|c|c|c|}
    \hline
    Atom & Li & Be & B   & C   & N   & O   & F   & Ne \\
    \hline
    PT Error & 5.6\% & 6.2\% & 7.8\% & 9.0\% & 10.0\% & 11.2\% &
    12.2\% & 13.0\% \\
    CI Error & 0.9\%& 0.6\%& 0.7\% & 0.8\% & 0.8\% & 1.0\% & 1.0\% & 1.1\%\\
    \hline
  \end{tabular}
  } 
  \caption{Percentage error in the ground state energy for the PT and minimal asymptotics-based CI method.}
  \label{Tab:GSPercentError}
 \end{center}
\end{table}

\vspace*{-2mm}
\noindent
Some insight can be
gained from comparing the CI orbitals resulting from energy minimization with the ``bare'' PT orbitals.
It is clear from Tables \ref{Tab:EnergyComparison1} and
\ref{Tab:EnergyComparison2} that $Z_1\approx Z-0.3$ and hence the PT
model $1s$ orbitals are a fair approximation to those in the CI
model. But this is not true for the $2s$ and $2p$ orbitals since
$Z_2$ is lower than $Z$ by about $2$, and $Z_3$ is lower by about
$2$ to $4$.

Physically this is intuitive from the idea that the 1s orbitals
partially screen the nuclear charge felt by the 2s and 2p orbitals,
making the 2s and 2p electrons behave as they would in the potential
of a nucleus with reduced nuclear charge.

Mathematically, one can at least explain {\it why} the $Z_i$ differ from their PT value of $Z$. The CI wavefunctions
satisfy the virial theorem (see Section \ref{sec:Asy}); by contrast the
deviation of the PT wavefunctions from the correct virial ratio between potential to kinetic energy of $-2$ is large,
because these states, being ground states of a non-interacting Hamiltonian, have a ratio of $-2$ for
(potential energy without electron repulsion) to kinetic energy. (From \cite{PT}, the actual virial ratios of the PT ground states
for Li, Be, B, C, N, O, F, Ne are -1.6969, -1.6881, -1.6615, -1.6379, -1.6173, -1.5956, -1.5778, -1.5615.)
\\[2mm]
We now discuss the obtained wavefunctions. Our work provides for the first time few-parameter,
explicit, closed-form wavefunctions for the low-lying
eigenstates of the atoms Li, Be, B, C, N, O, F, Ne which are of chemically relevant accuracy. These
can be used as a source of numerous theoretical insights.
\\[2mm]
As an important application, the wavefunctions given by Tables \ref{Tab:VeeMatrix1}--\ref{Tab:VeeMatrix2} and eq. (\ref{degestates})
and their ordering given in Tables \ref{Tab:EnergyComparison1}--\ref{Tab:EnergyComparison2}
confirm and make quantitative the qualitative mathematical picture of the periodic table obtained in
\cite{PT} by asymptotic analysis of the Schr\"odinger equation for strongly positive ions.
For instance, they affirm the conclusion of \cite{PT} that the empirical shell ordering rule of quantum chemistry
(as the electron number increases, the 2s shell is ``filled'' before the 2p shell) is only correct in a probabilistic sense.
In degenerate symmetry subspaces, the minimal CI eigenstates contain the two configurations
$1s^22s^22p^{N-4}$ and $1s^22p^{N-2}$ (see the discussion of 2s--2p resonance in \cite{PT}). The state with lower energy
is dominated by the first configuration, i.e. the coefficient for the part of
the wavefunction in $1s^22s^22p^{N-4}$ is larger than the part in $1s^22p^{N-2}$. The reverse is true for the higher
energy state. Nevertheless, the minority contributions are of significant size (36$\%$, 27$\%$ and 17$\%$ in case of the
B, Be, C ground state).

\subsection{Spectral gaps and ionization energies}

These are an extremely tough test of any model, due to the multiscale effect that they are smaller by two to three
orders of magnitude (see Table \ref{Tab:Multiscale}).

First, note how our eigenstate tables allow to trace spectral gaps to the size of individual Coulomb and exchange integrals,
revealing the cancellations that lead to the small size of gaps compared to total energies (see Table \ref{Tab:Multiscale}).

As an example of a $2s$--$2p$ spectral gap, consider the $^2S$ ground state and $^2P$ first excited state of Lithium.
Table \ref{Tab:VeeMatrix1} shows that the gap at fixed values of $Z_1$, $Z_2$, $Z_3$ is given by the difference in one-body energy and
interaction with the $1s$ shell of the $2p$ and $2s$ orbitals, $[(3|3)+2(11|33)-(13|31)] - [(2|2)+2(11|22)-(12|21)]$. Substituting for simplicity the bare
values $Z_i=Z=3$ into the table in Lemma \ref{L:VeeElements}, the difference between the Coulomb terms is only $\frac{16}{81}=0.19753...$ (and
that between the exchange terms only $\frac{32}{2187}=0.01463...$), which is
much smaller than the common part $2(1|1)+(2|2)+(11|11)=2(11)+(3|3)+(11|11)=-\frac{33}{4}=-8.25$ contained in each of the states.

As an example of an energy level splitting between two states with an equal number of
$1s$, $2s$ and $2p$ orbitals, consider the $^4S^o$ ground state and $^2D^o$ first
excited state of Nitrogen. A look at Table \ref{Tab:VeeMatrix2}
reveals that the energy difference consists only of
the exchange term $-3(34|43)$, which is present in the ground state
due to the parallel spins of the three $p$-orbitals, but absent in
the excited state.

Next, as shown in Tables \ref{Tab:EnergyComparison1} and
\ref{Tab:EnergyComparison2}, the spectral gaps for the CI
model are in good agreement with experimental data (most are
within $10\%$) and comparable to the predictions of numerical studies with a much larger number of degrees
of freedom \cite{Tatewaki94}.
Considering for example the first three spectral gaps of Nitrogen, Carbon and Oxygen, CI has the more accurate value in five
out of nine cases, and the less accurate value in the remaining four cases.

To achieve this accuracy, the minimal form (C') of relaxation of orbitals in the CI model is needed, as the ``bare'' PT orbitals, despite sharing
asymptotic exactness in the large nuclear charge limit, give very poor spectral gaps, with errors in the order of $200-300\%$.

\begin{figure}
\begin{center}
\resizebox{0.5\textwidth}{!}{
\includegraphics{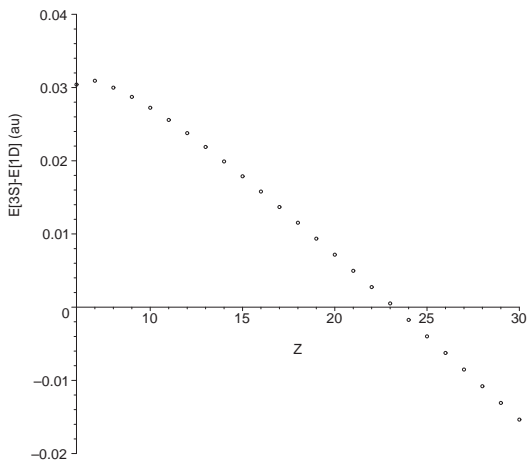}
}
\end{center}
\vspace{-2mm}
\caption{Minimal CI energy differences $E_{^3S^o}-E_{^1D^o}$ for the Carbon isoelectronic sequence
$N=6$, $Z=6, \dots, 28$. The predicted existence of a level crossing is confirmed by experiment \cite{NIST}.}
\label{Fig:crossing}
\end{figure}

One interesting qualitatively new feature introduced by the CI model is the possibility for energy levels to cross as the nuclear charge $Z$ varies
(see Figure \ref{Fig:crossing}) . This is due to the non-linearity of the energy levels in $Z$ arising from the minimization
over the dilation parameters $\undZ=(Z_1,Z_2,Z_3)$. (Note that for $Z_1=Z_2=Z_3=Z$,
the energy levels have the special form $-aZ^2+b_iZ$ \cite{PT}, yielding linearity of gaps in $Z$.)
This enables us to discuss, for example, the $^3S^o$ and $^1D^o$ states of the Carbon isoelectronic sequence.
We recall from \cite{PT} that both Hund's rules and the Hartree-Fock picture predict the universal ordering $E_{^3S^o}<E_{^1D^o}$,
which agrees with the experimental orderings for Carbon.
However, for $Z\ge 20$ the experimental ordering is found to be reversed.
This crossing is beautifully captured by the minimal CI model, this time for $Z\ge 23$.

We now discuss another important class of energy differences, ionization energies.
The latter are defined to be (writing $E_1(N,Z)$ to indicate the dependence of the ground state energy
on the number of electrons and the nuclear charge) $I(N,Z) = E_1(N-1,Z) - E_1(N,Z)$.
Physically this corresponds to the energy required to remove one electron from a system with nuclear charge $Z$ and $N$ electrons.
The calculated first ionization energies $I(N,N)$ of the minimal CI
Model, in atomic units, are as follows: He 0.8477, Li 0.1912, Be 0.3237, B 0.2346, C 0.3142, N 0.3960, O 0.2708, F 0.3958, Ne 0.4141.
The experimental ionization energies \cite{Huheey93} are: He 0.9036, Li 0.1980, Be 0.3426, B 0.3049, C 0.4138, N 0.5341, O 0.5000, F 0.6402, Ne 0.7925.

\begin{figure}
 \begin{center}
  \resizebox{0.4\textwidth}{!}{
   \includegraphics{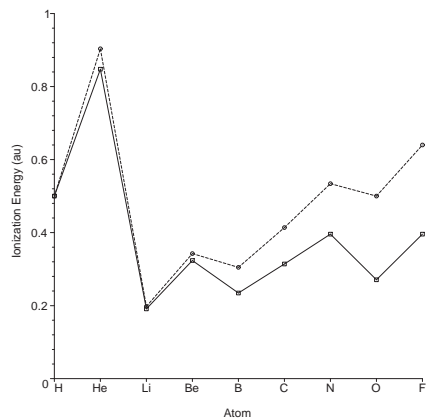}
  } 
 \end{center}
 \vspace{-2mm}
 \caption{Experimental (dashed) and minimal CI (solid) ionization energies.}
 \label{Fig:IonizationEnergies}
\end{figure}

Figure \ref{Fig:IonizationEnergies} shows that the qualitative
prediction for the ionization energies is very good when compared to
experimental data. In particular, all local minimizers (H, Li, B, O), local maximizers (He, Be, N), global minimizers (Li) and global maximizers (He)
are predicted correctly. This is all the more remarkable when remembering that tiny eigenvalue differences for partial differential
operators on very high-dimensional spaces up to $\R^{27}$ are under consideration here.

Quantitatively, for the smaller atoms our results are comparable to (and in case of Be better than)
MDHF calculations with much larger basis sets up to $20s11p$ \cite{Jorge01}. For the larger atoms the minimal dimensionality of
our CI subspace finally makes itself felt, and a larger subspace (e.g. as described at the end of Section 2.4)
would be needed to make the qualitative agreement quantitative.

Again, it is also instructive to compare with the PT model \cite{PT}. Its ionization energies, which are easily read off from the exact results of \cite{PT},
even turn out to have the wrong sign. This shows that relaxation of orbitals is important for the description of ionization processes,
and that the relaxation step (C) in the minimal asymptotics-based CI model is essential for understanding the nontrivial experimental graph in Figure \ref{Fig:IonizationEnergies}.
\vspace*{4mm}

\noindent
{\bf Acknowledgements} The research of B.G. was supported by a graduate scholarship from EPSRC. We thank
P.Gill for helpful comments.


\bibliographystyle{alpha}
\bibliography{CI}

\noindent
Address of authors: \\[2mm]
Gero Friesecke \\
Center for Mathematics, TU Munich, Germany, {\tt gf@ma.tum.de} \\[2mm]
Benjamin D. Goddard \\
Mathematics Institute, University of Warwick, U.K. {\tt b.d.goddard@warwick.ac.uk}

\end{document}